\documentclass[10pt,aps,twocolumn,showpacs,showkeys,floatfix]{revtex4}
\usepackage[dvips]{graphicx}

\begin{document}
\bibliographystyle{plain}
\title{Bistable Chimera Attractors on a Triangular Network of Oscillator Populations\\ \small \today}
\author{\firstname Erik A. \surname Martens}
\affiliation{
  Max Planck Research Group for Biological Physics and Evolutionary Dynamics, 
  Max Planck Institute for Dynamics and Selforganization (MPIDS), G\"{o}ttingen 37073, Germany
}
\email{erik.martens@ds.mpg.de}

\begin{abstract}
We study a triangular network of three populations of coupled phase oscillators with identical frequencies. The populations interact nonlocally, in the sense that all oscillators are coupled to one another, but more weakly to those in neighboring populations than to those in their own population. This triangular network is the simplest discretization of a continuous ring of oscillators. 
Yet it displays an unexpectedly different behavior: in contrast to the lone stable chimera observed in continuous rings of oscillators, we find that this system exhibits \emph{two coexisting stable chimeras}.
Both chimeras are, as usual, born through a saddle node  bifurcation. As the coupling becomes increasingly local in nature they lose stability through a Hopf bifurcation, giving rise to breathing chimeras, which in turn get destroyed through a homoclinic bifurcation. Remarkably, one of the chimeras reemerges by a reversal of this scenario as we further increase the locality of the coupling, until it is annihilated through another saddle node bifurcation.
\end{abstract}
\maketitle
\section{Introduction}\label{sec:introduction}
While studying a continuum of identical oscillators on a ring with nonlocal coupling, Kuramoto \emph{et al.} \cite{battogtokh2002cin} discovered a remarkable state where the population of oscillators splits into two subpopulations, where one is synchronized and the other is desynchronized. This state was later dubbed a \emph{chimera}, alluding to a monster in Greek mythology that consists of incongruous parts. Since then, several groups have explored the nonlinear dynamics of chimera states \cite{battogtokh2002cin,abrams2004cro,abrams2004csc,shima2004rsw,kawamura2007ciw,kawamura2007hsn,abrams2008smc,laing2009csh,omelchenko2008cnl}. Their emergence on the ring was first investigated analytically by Abrams and Strogatz \cite{abrams2004cro,abrams2004csc}. They found that chimera states were born through a saddle node bifurcation, which appears to be the typical scenario for the emergence of chimeras on all network topologies investigated so far. Shima and Kuramoto  \cite{shima2004rsw} showed that chimeras also exist on 2D lattices with free boundaries in the shape of spiral waves:  here, the center of the spiral, characterized by a topological defect, is replaced by a desynchronized core with finite positive radius \cite{martens2010swc}. 

A recent breakthrough in the field of coupled oscillator systems was made by Ott and Antonsen who showed that the dynamics of infinitely many oscillators could be reduced to a system of finite ordinary differential equations.
The method has since then been generalized and put into a formal mathematical context by Pikovsky and Rosenblum \cite{pikovsky2008pid} and Marvel \emph{et al.} \cite{mirollo2009} independently, who show that the dynamics of each population may be reduced to a flow described by three variables plus constants of motion. Abrams \emph{et al.} \cite{abrams2008smc} employed the method to investigate a system of two interacting populations of phase oscillators;
they were able to calculate the saddle node bifurcation for this system analytically and showed that the chimera states undergo a further change of stability to become \emph{breathing chimeras} via a supercritical Hopf bifurcation. 

Whereas many of these studies have been focusing on the nature of the chimera in idealized settings, others have been studying how chimeras would occur or behave in systems closer to real world situations: Omel'chenko \emph{et al.}  \cite{omelchenko2008cnl} show that a network of globally coupled oscillators, subjected to delayed feedback stimulation with spatially decaying profile, also induces chimera states; thereby they argue that chimera states indeed are a generic feature of coupled oscillator systems. Delay coupled systems are also investigated by Sethia \emph{et al.} \cite{sethia2005ccs} who discover the clustering of chimera states. 
Makovetskiy \emph{et al.} \cite{makovetskiy2008rsw,makovetskiy2005} mention they found chimera-like states in systems of three level cellular automata related to lasing systems in combination with spiral waves.  
Laing investigates the robustness of chimera states against the introduction of nonidentical oscillators with heterogeneous frequencies in several systems \cite{laing2009hkn}.
For the case of two nonlocally coupled populations studied by Abrams \emph{et al.} \cite{abrams2008smc}, Laing shows how chimeras may both be destabilized and stabilized as the strength of heterogeneity (i.e. the width of the frequency distribution) of the oscillators is varied\cite{laing2009csh}. Finally, we mention that in neuroscience, spatially localized ``bumps'' of neural activity are found in networks of spiking neurons - such states have been proposed as mechanisms for visual orientation tuning and working memory, and have been related to chimeras \cite{laing2001sbn,laing2001nis}.

Important questions still remain. In particular, which topologies allow for chimera states? In other terms, can we classify the network structures that allow for chimeras?  Even on one-dimensional domains the situation is unclear. It has been shown that chimeras may exist on a ring \cite{battogtokh2002cin} with a continuum of oscillators, but 
do chimeras also exist on a finite line segment or the infinite line, i.e. on a non-periodic domain? 
The observation that chimeras exist in the shape of spiral waves on 2D domains \cite{shima2004rsw,martens2010swc} is a hint that this might be possible.

One may seek to answer this particular question by lowering the dimensionality of the problem and discretizing a one-dimensional domain using oscillators populations. As in Abrams \emph{et al.} \cite{abrams2008smc}, one assumes that oscillators within a given population are coupled more strongly to each other than they are to those in neighboring populations, thereby defining a spatial structure on the network. Thus the simplest network that exhibits both chain-like and ring-like character consists of three populations, as shown in Fig. \ref{fig:networkstriangle}. By tuning a new structural parameter $c$, a rotationally symmetric (ring-like or ``triangular'') network (Fig. \ref{fig:networkstriangle} (b)) is deformed into a less symmetric, chain-like structure (Fig. \ref{fig:networkstriangle} (a)). The impact of changing the network topology in such ways is discussed in a companion article \cite{martens2009ntc}.

This paper is wholly devoted to the study of the purely triangular network as shown in Fig. \ref{fig:networkstriangle} (b). While examining this simple case, we surprisingly found the \emph{coexistence of two stable chimera attractors}. These attractors differ in the number of desynchronized populations (one or two). 
Their occurrence is unexpected because the ring with a continuum of oscillators \cite{abrams2004csc,abrams2004cro}, sharing the same rotational symmetry, has only a \emph{single} stable chimera state.
The finding of multiple coexisting chimera attractors is novel and is relevant in various contexts. First, if we let the number of oscillator populations go to infinity, we would retrieve the system of a ring studied by Abrams \emph{et al.}; however, this system is not known to yield any multistable chimeras, which forms an interesting discrepancy.
Second, the finding of bistable chimeras is noteworthy in the context of the question raised just recently \cite{motter2010} about what the basins of attraction leading to chimera and globally synchronized states actually are. Third, in connection with the above mentioned 'bump states', multistability of chimera states may be relevant in the study of mathematical neural models, where so called switching processes \cite{aswhinnature2005} and the competition of attractors in working memory \cite{hoppensteadt1999} are of interest (see Discussion). 

The organization of the article is as follows. We introduce the governing equations in Section II and summarize the derivation to obtain the reduced set of equations, as described in \cite{abrams2008smc}. Next we derive the equations implied by special symmetries consistent with chimera states. These are analyzed in Section III, together with all their possible bifurcations. Section IV discusses our results in the context of numerical simulations. Finally, our findings are discussed in Section V. Additional results on stability of some simple symmetric states for arbitrary networks are outlined in the Appendix.

\section{Governing equations}\label{sec:governingequations}
The governing equations are given by
\begin{eqnarray}\label{eq:goveqn}
	\frac{d}{dt}\theta_i^{\sigma}&=&\omega + \sum_{\sigma'=1}^3 \frac{K_{\sigma\sigma'}}{N_{\sigma'}}\sum_{j=1}^{N_{\sigma'}} \sin{(\theta_j^{\sigma'} -\theta_i^{\sigma}  - \alpha)},
\end{eqnarray}
where the phases of the oscillators are defined by $\theta$, $i$ denotes the individual oscillators belonging to the population with indices $\sigma = 1,2,3$, each of which has $N_{\sigma}$ oscillators, and where $\alpha$ is the phase lag parameter.

\begin{figure}[ht]
	\begin{center}
		\includegraphics[width=.32\textwidth]{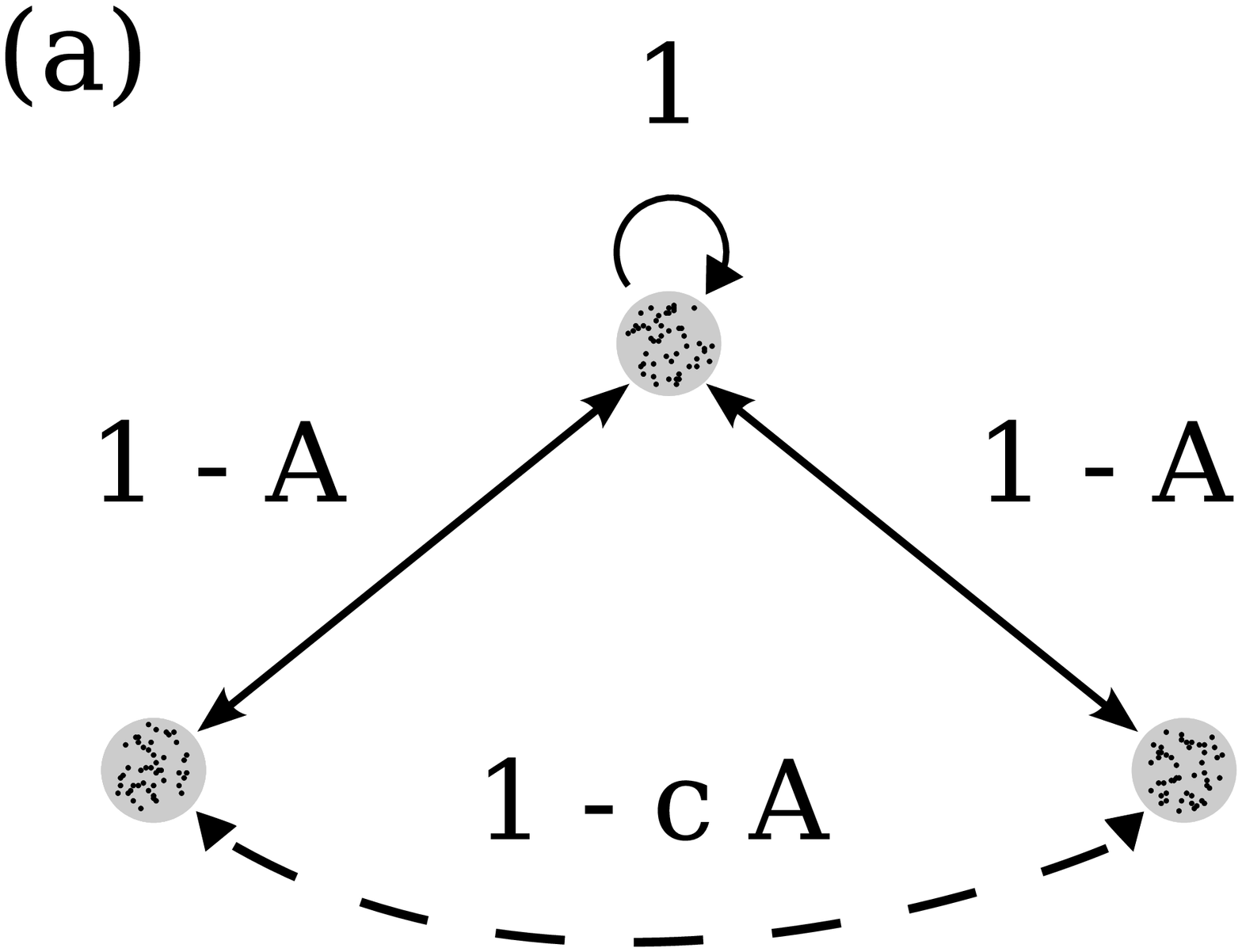}
		\includegraphics[width=.32\textwidth]{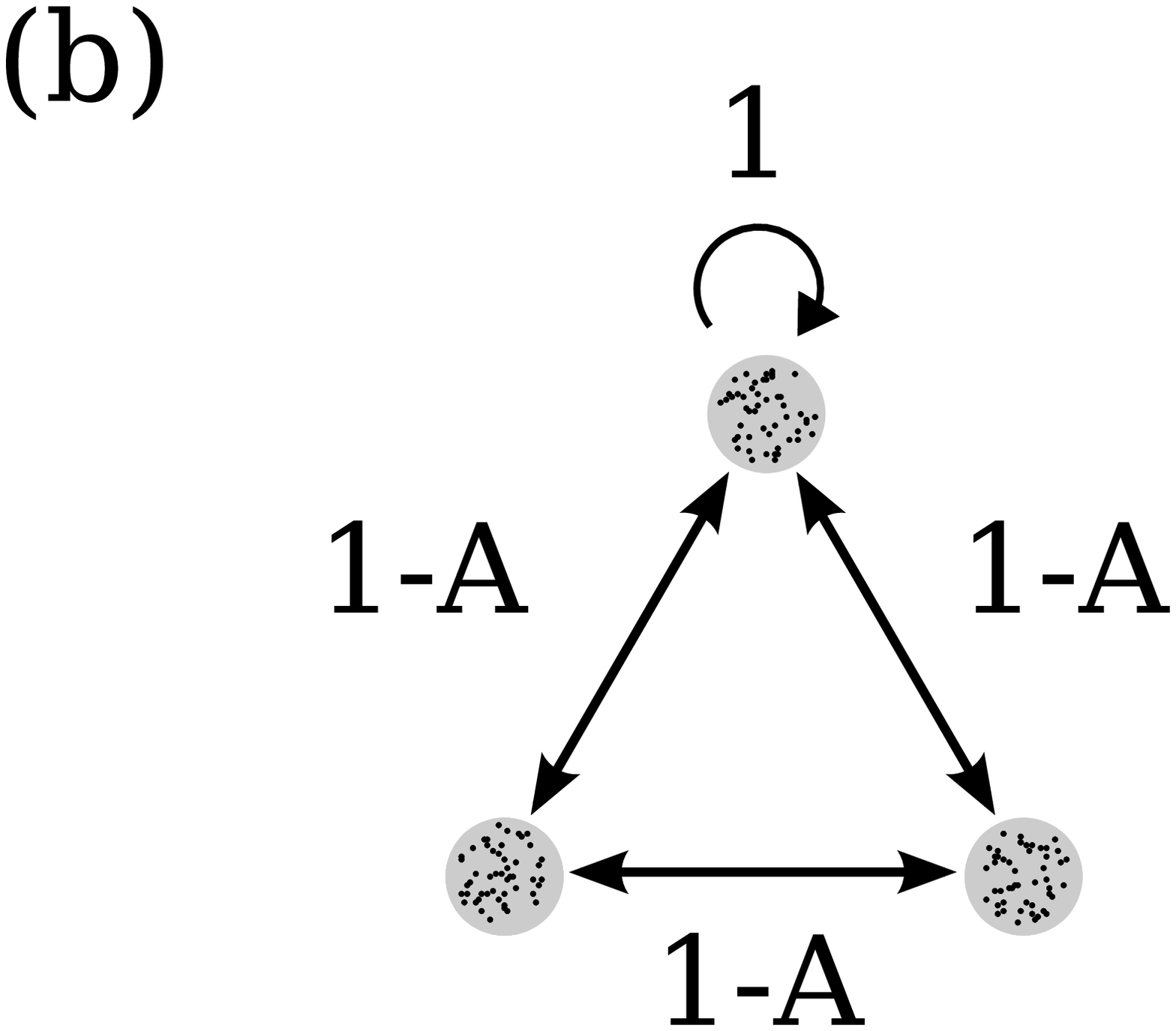} 
	\end{center}
\caption {Networks of three populations of oscillators. The gray disks symbolize the populations of oscillators, populated by individual oscillators symbolized by black dots. Their bidirectional coupling is represented by black arrows. (a) Chain-like general case with structural tuning parameter $c$. (b) Triangular network structure, corresponding to $c=1$. Each population has a self-coupling of unit strength $1$, and is coupled to the neighboring populations with strength $1-A$. }
\label{fig:networkstriangle}
\end{figure}
The coupling kernel $K_{\sigma\sigma'}$ describes the strength between populations $\sigma$ and $\sigma'$. The coupling strength is assumed to decay with increasing separation between the populations on the network. Within a population, the oscillators interact with strength $K_{\sigma\sigma'}=1$. Neighboring populations couple more weakly, with strength $1-A$ as displayed in Fig. \ref{fig:networkstriangle} (b). We then have
\begin{equation}\label{eq:kernel}
 K_{\sigma\sigma'} = \left(\begin{array}{lcr}
			1    & 1-A & 1-A\\
                        1-A  & 1   & 1-A\\
			1-A & 1-A & 1\
                       \end{array}
\right). 
\end{equation}
In the case of $A=0$, we retrieve the case of a globally coupled network. Thus $A$ quantifies how 'far' we are from global coupling.  This network has the same rotational symmetry as a continuum of oscillators on a ring, studied by \cite{battogtokh2002cin,abrams2004csc,abrams2004cro}, and generalizes the problem with two populations discussed by Abrams \emph{et al.} \cite{abrams2008smc}. 

\subsection{Reduction to low-dimensional system}
We consider the  limit of infinitely large populations, i.e. $N_{\sigma}\rightarrow \infty$. This allows us to reduce the problem to a finite set of equations, using the ansatz introduced by Ott and Antonsen \cite{ott2008ldb}, outlined below. In this limit, it is natural to describe the dynamics of the system in terms of the oscillator density distribution $f^{\sigma}(\theta)$, which evolves according to the continuity equation
\begin{eqnarray}\label{eq:continuityeqn_triangular}
 	\frac{\partial f^{\sigma}}{\partial t} + \frac{\partial}{\partial \theta}(f^{\sigma}v^{\sigma}) &=& 0.\
\end{eqnarray}
The velocity of the oscillators is then given by
\begin{eqnarray}\nonumber
 	v^{\sigma}&=&\omega + \sum_{\sigma'=1}^3 K_{\sigma\sigma'}\int_0^{2\pi} \sin{(\theta'-\theta-\alpha)}f^{\sigma'}(\theta',t)d\theta'. \\
\end{eqnarray}
To keep the notation simple we denote $\theta_{\sigma}$ by $\theta$ and $\theta_{\sigma'}$ by $\theta'$. It proves convenient to define a complex order parameter
\begin{eqnarray}
 	z^{\sigma}(t) &=& \sum_{\sigma'=1}^3 K_{\sigma\sigma'}\int_0^{2\pi}e^{i\theta'}f(\theta',t)d\theta',\
\end{eqnarray}
which defines a weighted average over all oscillators; we therefore refer to this as the \emph{global order parameter}.
Following Kuramoto's footsteps \cite{kuramoto1984owt,kuramoto2006mft,battogtokh2002cin}, we rewrite the velocity in terms of the order parameter and find
\begin{eqnarray}\nonumber
 	v^{\sigma}&=&\omega + \textrm{Im}\left[ e^{-i\theta} e^{-i\alpha} \sum_{\sigma'=1}^3 K_{\sigma\sigma'}\int_0^{2\pi}e^{i\theta'}f(\theta',t)d\theta' \right]\\
		  &=&  \omega + \frac{1}{2i}\left[ e^{-i\theta} e^{-i\alpha} z_{\sigma}(t) - e^{i\theta} e^{i\alpha} z_{\sigma}^*(t) \right].\
\end{eqnarray}
Following Ott and Antonsen \cite{ott2008ldb}, we now restrict attention to a special class of density functions in the form of a Poisson kernel
\begin{eqnarray}\label{eq:poissonkernel}
 	f^{\sigma}(\theta,t)&=& \frac{1}{2\pi}\left\{1+\left[\sum_{k= 1}^{\infty}(a_{\sigma}(t)e^{i\theta})^k + c.c.\right]\right\},\
\end{eqnarray}
where c.c. is the complex conjugate of the expression under the sum. 
The implications of this special ansatz and its validity will be explained in more detail in Section \ref{sec:discussion}.
Substitution of $f_{\sigma}$ and $v_{\sigma}$ into the continuity equation (\ref{eq:continuityeqn_triangular}) yields an exact solution, so long as $a_{\sigma}$ evolves according to
\begin{eqnarray}\label{eq:amplequations}
 	0&=& \dot{a}_{\sigma}\ + i \omega a_{\sigma}+\frac{1}{2}a_{\sigma}^2\, z_{\sigma}\, e^{-i\alpha} -\frac{1}{2}\,z^*_{\sigma}\,e^{i\alpha}.\
\end{eqnarray}
It remains to express the order parameter in terms of this ansatz. We find
\begin{eqnarray}
 	z_{\sigma}(t)&=& \sum_{\sigma'=1}^3 K_{\sigma\sigma'}a_{\sigma'}^*(t).\
\end{eqnarray}
Finally, we express the amplitude $a_{\sigma}$ in polar coordinates as
\begin{eqnarray*}
	a_{\sigma}=\rho_{\sigma}e^{-i\phi_{\sigma}}.\
\end{eqnarray*}
By division of (\ref{eq:amplequations}) by $e^{-i\phi_{\sigma}}$, we obtain
\begin{eqnarray*}
	0 &=& \dot{\rho_{\sigma}} - i \dot{\phi_{\sigma}}\rho_{\sigma} + i\omega\rho_{\sigma} \\
	&&+ \frac{1}{2} \rho_{\sigma}^2\sum_{\sigma'=1}^3K_{\sigma\sigma'}\rho_{\sigma'}e^{i(\phi_{\sigma'}-\phi_{\sigma}-\alpha)}\\
	&&- \frac{1}{2}\sum_{\sigma'=1}^3 K_{\sigma\sigma'}\rho_{\sigma'}e^{-i(\phi_{\sigma'}-\phi_{\sigma}-\alpha)},\
\end{eqnarray*}
and by separation of real and imaginary parts we find 
\begin{eqnarray}\label{eq:lodimbeta}\nonumber
 \dot \rho_{\sigma}&=& \frac{1-\rho_{\sigma}^2}{2}\sum_{\sigma'=1}^{3}K_{\sigma\sigma'}\rho_{\sigma'}\sin{(\phi_{\sigma'}-\phi_{\sigma}+\beta)}, \\\nonumber
\dot \phi_{\sigma}&=&\omega- \frac{1+\rho_{\sigma}^2}{2\rho_{\sigma}}\sum_{\sigma'=1}^{3}K_{\sigma\sigma'}\rho_{\sigma'}\cos{(\phi_{\sigma'}-\phi_{\sigma}+\beta)},\\
\end{eqnarray}
where we introduce the definition 
\begin{eqnarray}
	\beta=\pi/2-\alpha.\
\end{eqnarray}
These equations describe the dynamics of our system in terms of the variables $a_{\sigma}$. Notice that in contrast to $z_{\sigma}$, they do not represent averages over all populations, and therefore, we refer to them as \emph{local order parameters}. Thus any synchronized population $\sigma$ of oscillators is characterized by $\rho_{\sigma}=1$. (These results are trivially generalized to the case of a network with arbitrarily many populations $\sigma = 1,2,\ldots,N$.)

In what follows, we will make particular assumptions about the symmetries of the solutions that we expect to find. Then we will analyze existence, stability and bifurcations of the states of interest.

\subsection{Manifold of symmetric states (SDS and DSD)}
To make progress on analyzing the chimera states, we will have to make certain symmetry assumptions. Perfectly synchronized populations have $\rho_{\sigma}=1$; desynchronized populations have $\rho_{\sigma}<1$, and consist of oscillators that drift relative to one another and to the synchronized populations.
Let $S$ and $D$ denote synchronized and desynchronized populations, respectively. 
Then in  a triangular network, we can distinguish only two chimera states, namely $SDS$ (sync-drift-sync) and $DSD$ (drift-sync-drift); all other permutations of $S$ and $D$ give equivalent states because of the rotational invariance inherent to the triangular network. 

Another class of solutions can be described as $SSS$, corresponding to a globally synchronized state where all oscillators are in sync, but with different synchronized mean phases $\phi_i$.
We may distinguish three cases, i.e. $\phi_1=\phi_2=\phi_3$,  $\phi_1=\phi_3\neq\phi_2$ and the state where all phase angles are different. These solutions are analyzed in Appendix A. These states are of less interest to us and we restrict ourselves to chimera states and their emergence in parameter space in the following.

The symmetry of our coupling kernel (\ref{eq:kernel}) in combination with $\rho_1 = \rho_3$ implies that $\phi_1=\phi_3$ 
and hence populations $1$ and $3$ are phase-locked. The $SDS$ state is then defined via 
\begin{eqnarray*}
  \rho_1 &=& \rho_3 = 1 \textrm{ and } \rho\equiv\rho_2<1,\\
  \phi_1&=&\phi_3,\
\end{eqnarray*}
whereas the  $DSD$ state is given by
\begin{eqnarray*}
  \rho&\equiv&\rho_1 = \rho_3<1 \textrm{ and } \rho_2=1.   \\
  \phi_1&=&\phi_3,\
\end{eqnarray*}
We define the phase difference of the angular order parameter between the synchronized and desynchronized states by
\begin{eqnarray}
	\psi=\phi_1-\phi_2=\phi_3-\phi_2.\
\end{eqnarray}
Applying these symmetry assumptions to (\ref{eq:lodimbeta}) and  substituting the coupling kernel defined in (\ref{eq:kernel}), we obtain the equations describing the $SDS$ states
\begin{eqnarray}\label{eq:SDShybrid}	\nonumber
	\dot \rho&=&  \frac{1-\rho^2}{2}\left[2(1-A)\sin{(\psi+\beta)}+\rho\sin{\beta}\right],\\\nonumber
	\dot \psi&=& -(2-A)\cos{\beta} - (1-A)\rho\cos{(-\psi+\beta)}\\
	&+&\frac{1+\rho^2}{2\rho}\left[2(1-A)\cos{(\psi+\beta)} +\rho\cos{\beta}\right],\
\end{eqnarray}
and the $DSD$ states
\begin{eqnarray}\label{eq:DSDhybrid}\nonumber
	\dot \rho&=&  \frac{1-\rho^2}{2}\left[(2-A)\rho\sin{\beta} + (1-A)\sin{(-\psi+\beta)}\right],\\\nonumber
	\dot \psi&=& -\frac{1+\rho^2}{2\rho}\left[(2-A)\rho\cos{\beta} + (1-A)\cos{(-\psi+\beta)}\right]\\     
	&+&2(1-A)\rho\cos{(\psi+\beta)} + \cos{\beta}.\
\end{eqnarray}
Note that these equations hold only if we restrict attention to symmetry-preserving perturbations. The fixed points of (\ref{eq:SDShybrid},\ref{eq:DSDhybrid}) correspond to phase-locked solutions of the original system (at the macroscopic level of the local order parameters).

By reduction of the full system of oscillators (\ref{eq:goveqn}) to a low dimensional system for the local order parameters, we have cast our problem into a two dimensional system represented by Eqs. (\ref{eq:SDShybrid},\ref{eq:DSDhybrid}). 
This enables us to study the problem in the phase plane.

\section{Analysis}\label{sec:analysis}
\subsection{Phase portraits}
\begin{figure*}[ht]
	\begin{center}
		\includegraphics[width=0.32\textwidth]{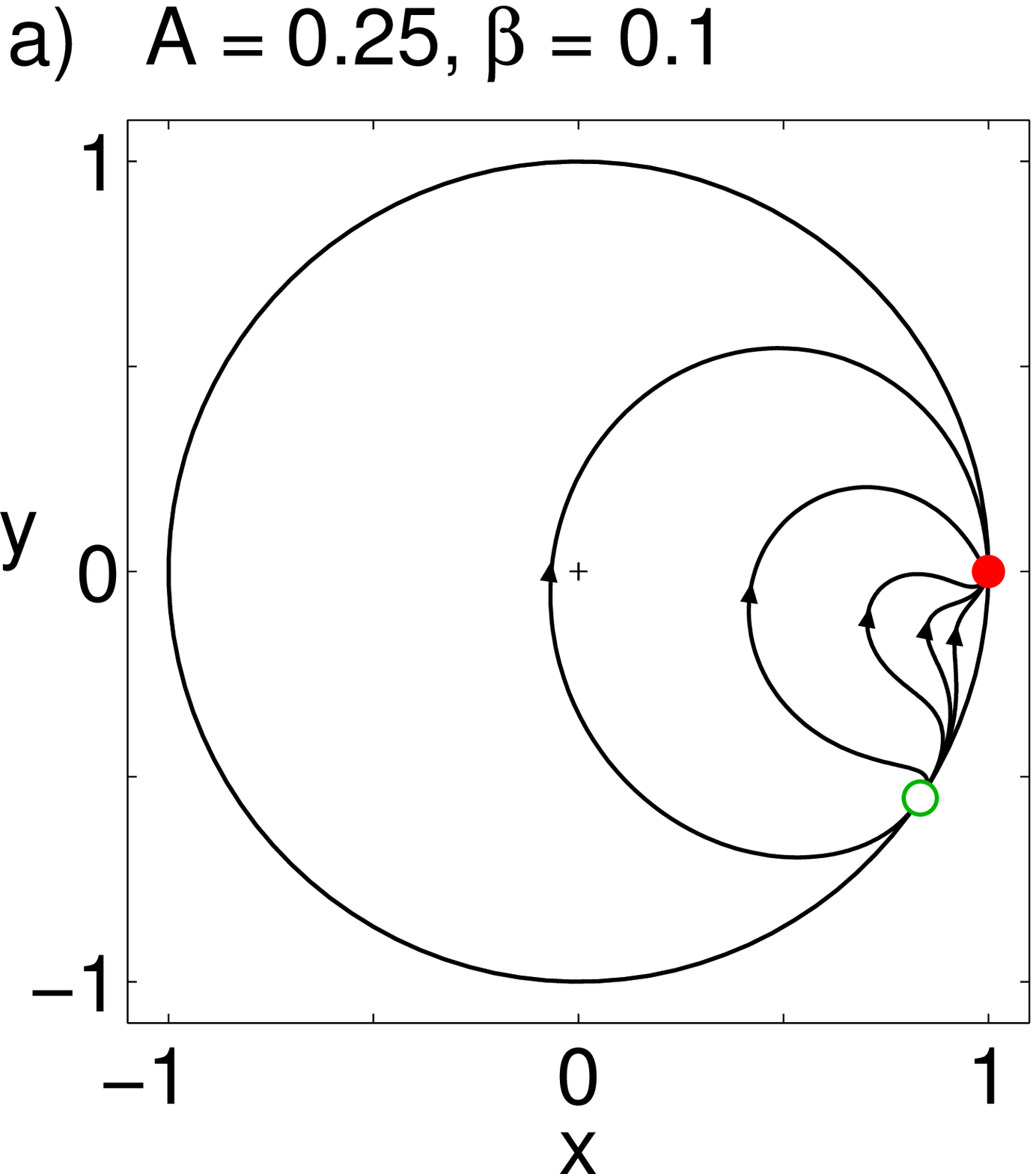} \hspace{0.5cm}
		\includegraphics[width=0.32\textwidth]{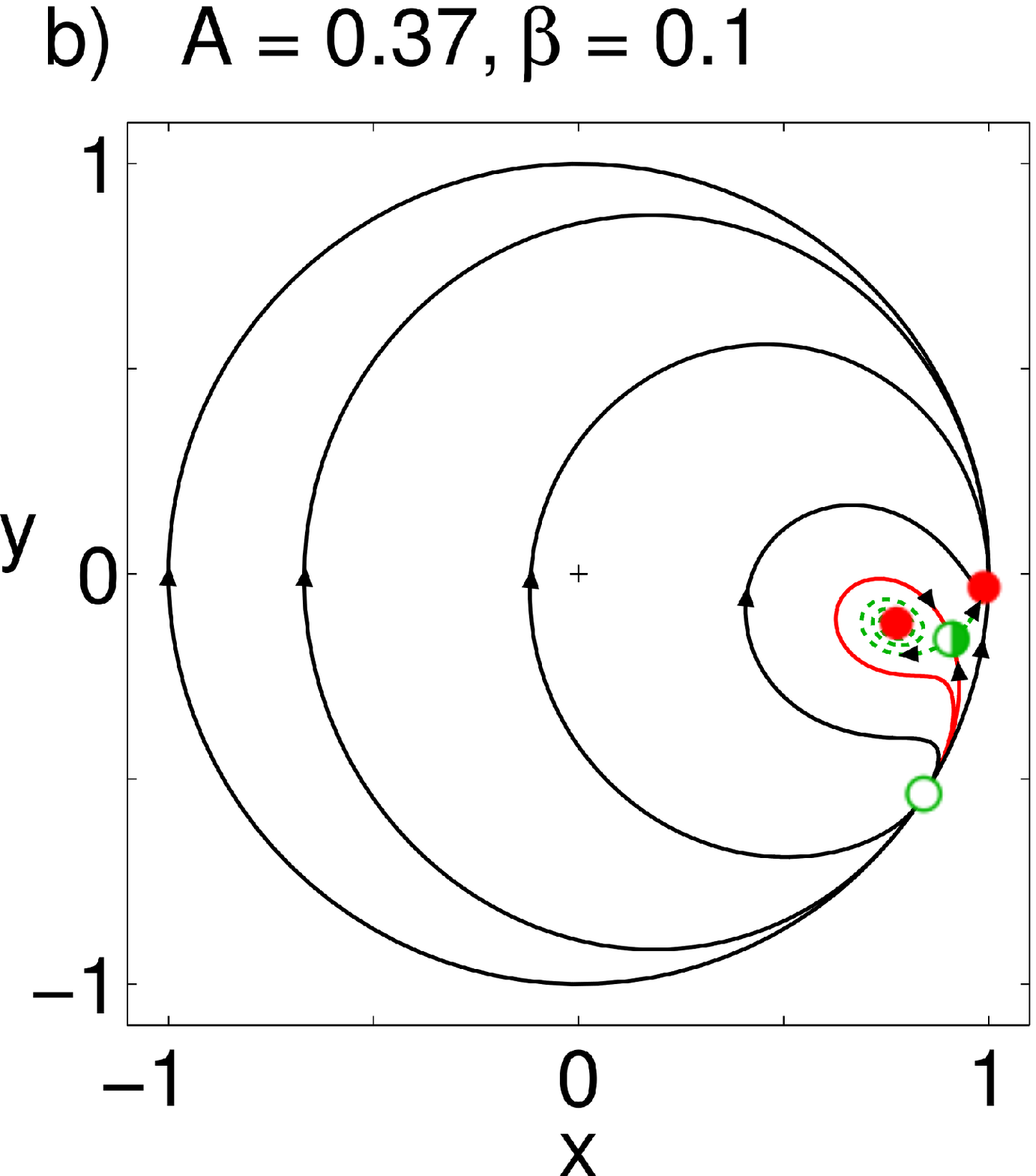}\\\vspace{0.3cm}
		\includegraphics[width=0.32\textwidth]{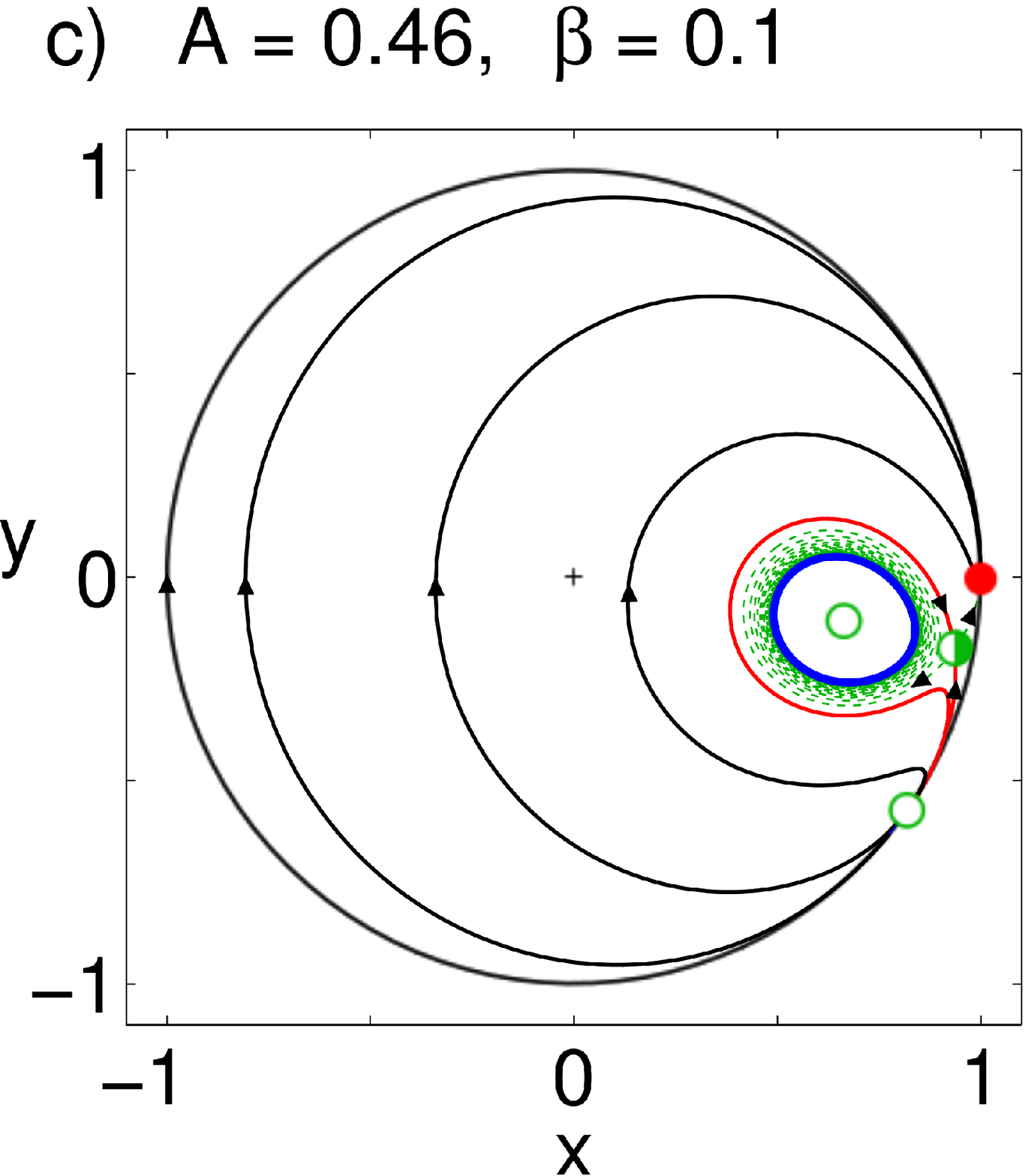} \hspace{0.5cm}
		\includegraphics[width=0.32\textwidth]{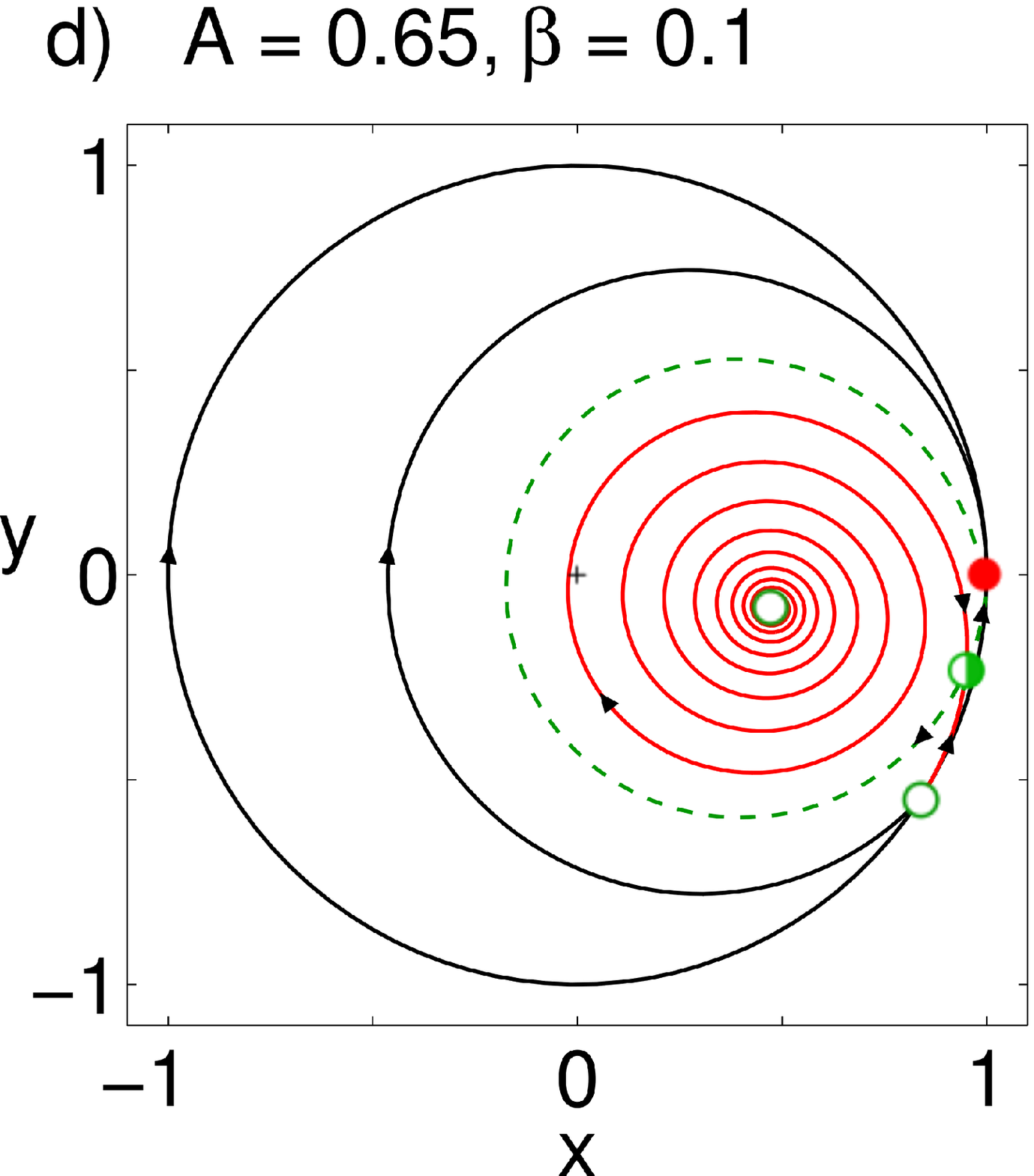} 		
	\end{center}
\caption{Phase portraits for the $SDS$ chimera, with increasing values of $A$ at constant $\beta$. Real ($x$) and imaginary ($y$) components of $\overline{a}_{(\sigma)}$ are shown. The unit circle is displayed in gray. Stable fixed points are shown as solid (red) and half-filled (green) circles, respectively. Limit cycles are emphasized in blue color. Stable and unstable manifolds are shown as red solid and green dashed trajectories, respectively. The point in $(\rho,\psi)=(1,0)$ is a nodal sink. The position of the nodal source depends on $\beta$ and moves in clockwise direction with growing values of $\beta$. }
\label{fig:pp_SDS_hybridc1}
\end{figure*}

\begin{figure*}[ht]
 	\begin{center}
		\includegraphics[width=0.32\textwidth]{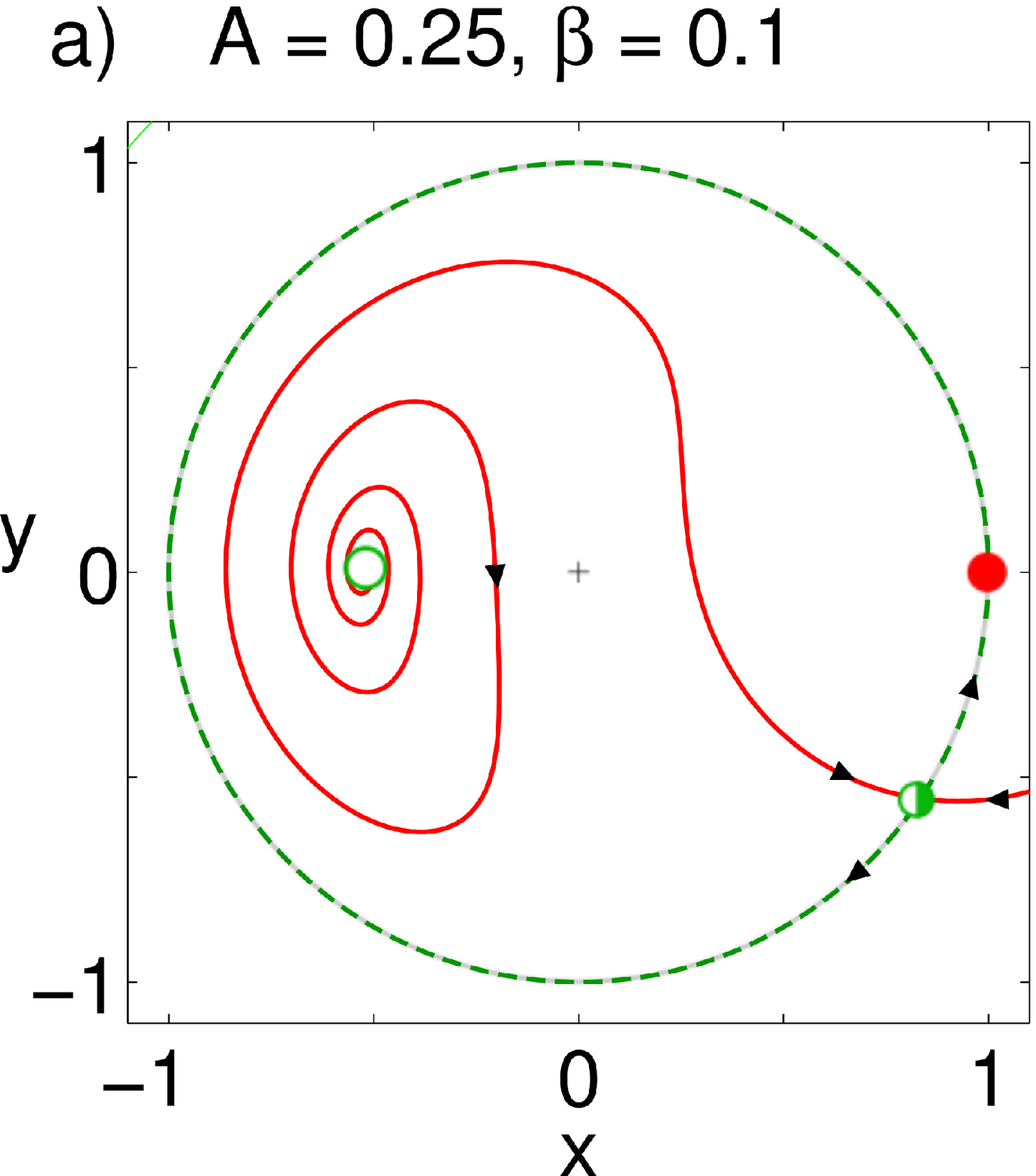} 
		\includegraphics[width=0.32\textwidth]{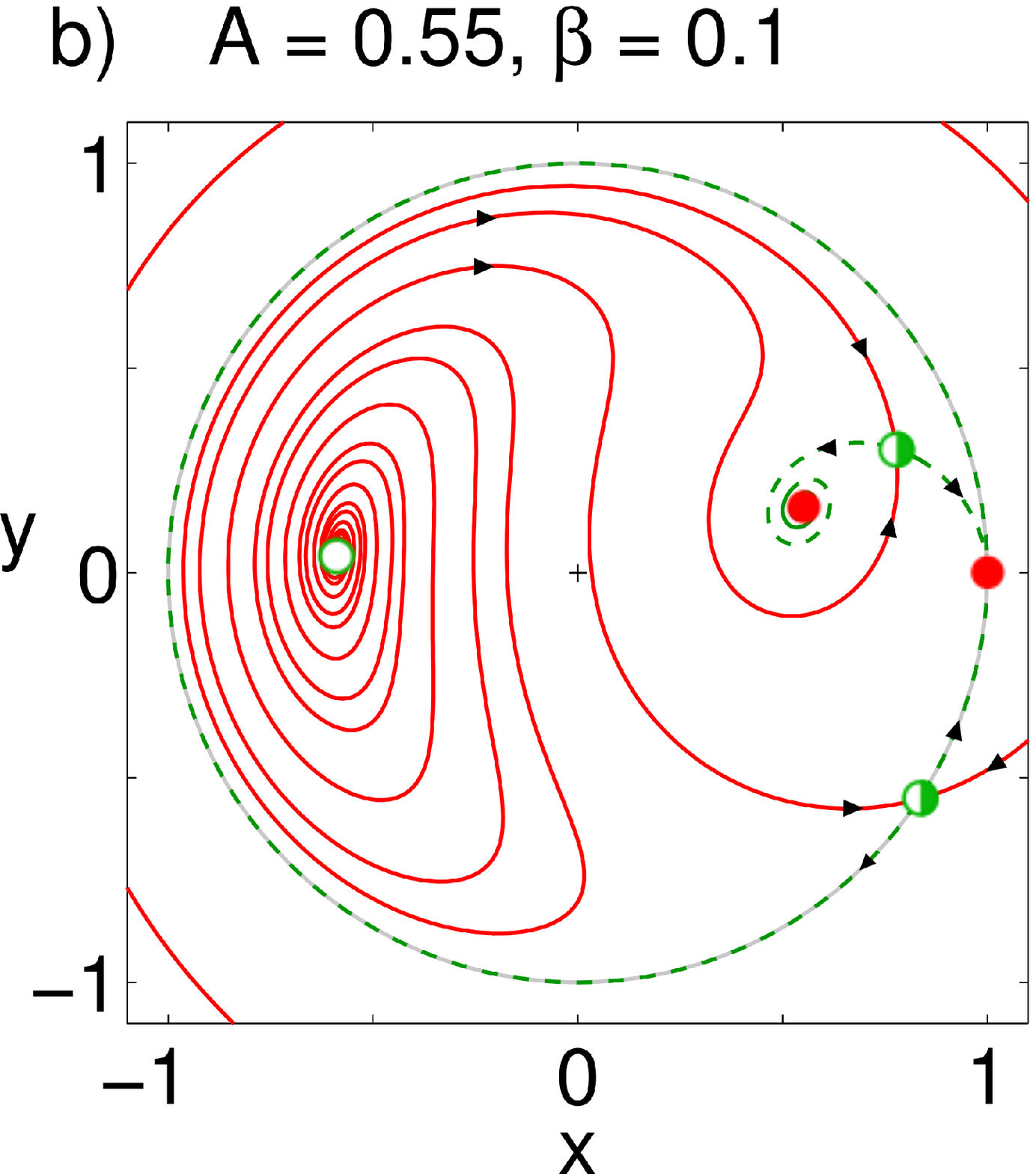} 		
		\includegraphics[width=0.32\textwidth]{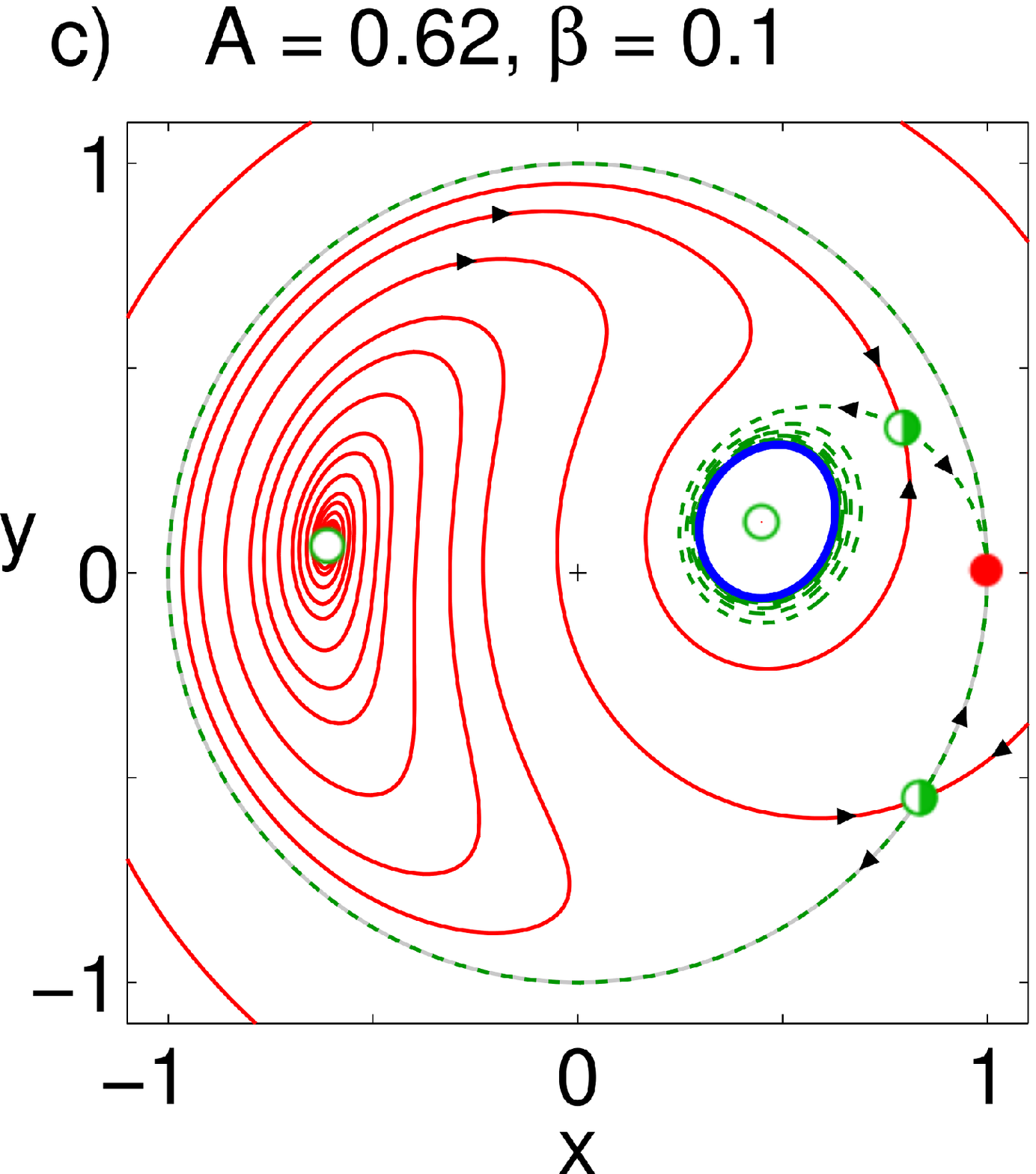} \\
		\vspace{.3cm}
		\includegraphics[width=0.32\textwidth]{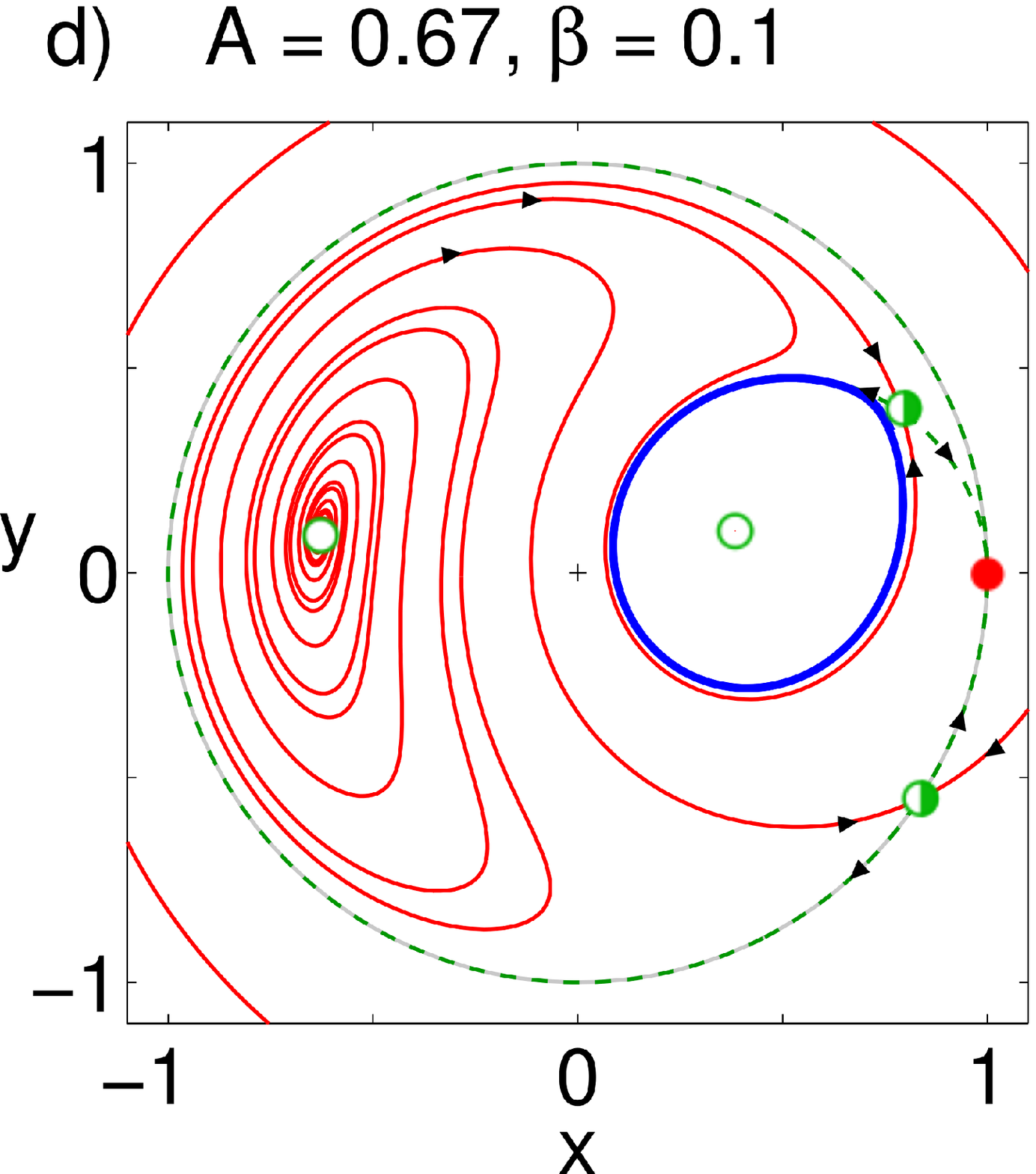} 
		\includegraphics[width=0.32\textwidth]{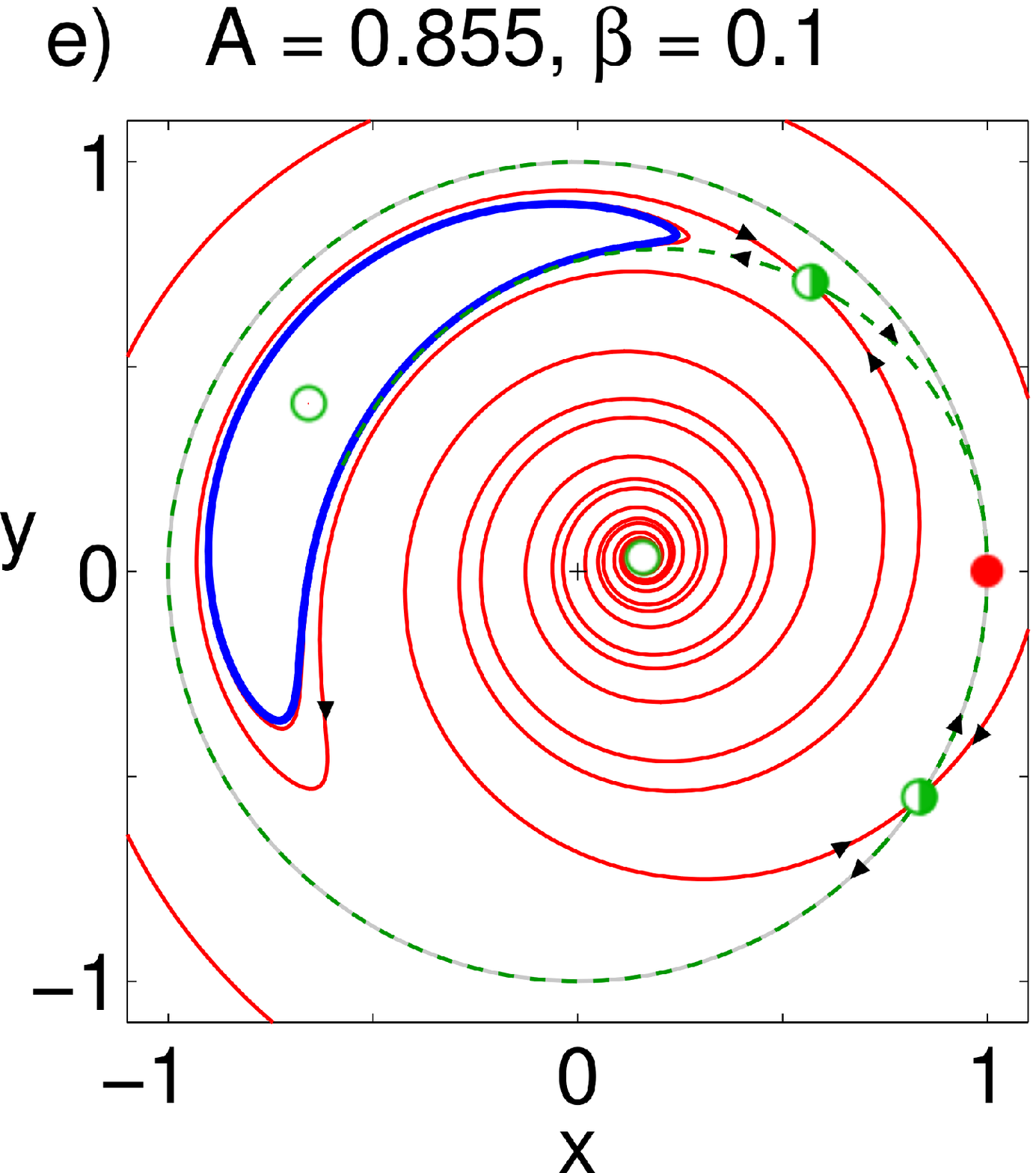} 		
		\includegraphics[width=0.32\textwidth]{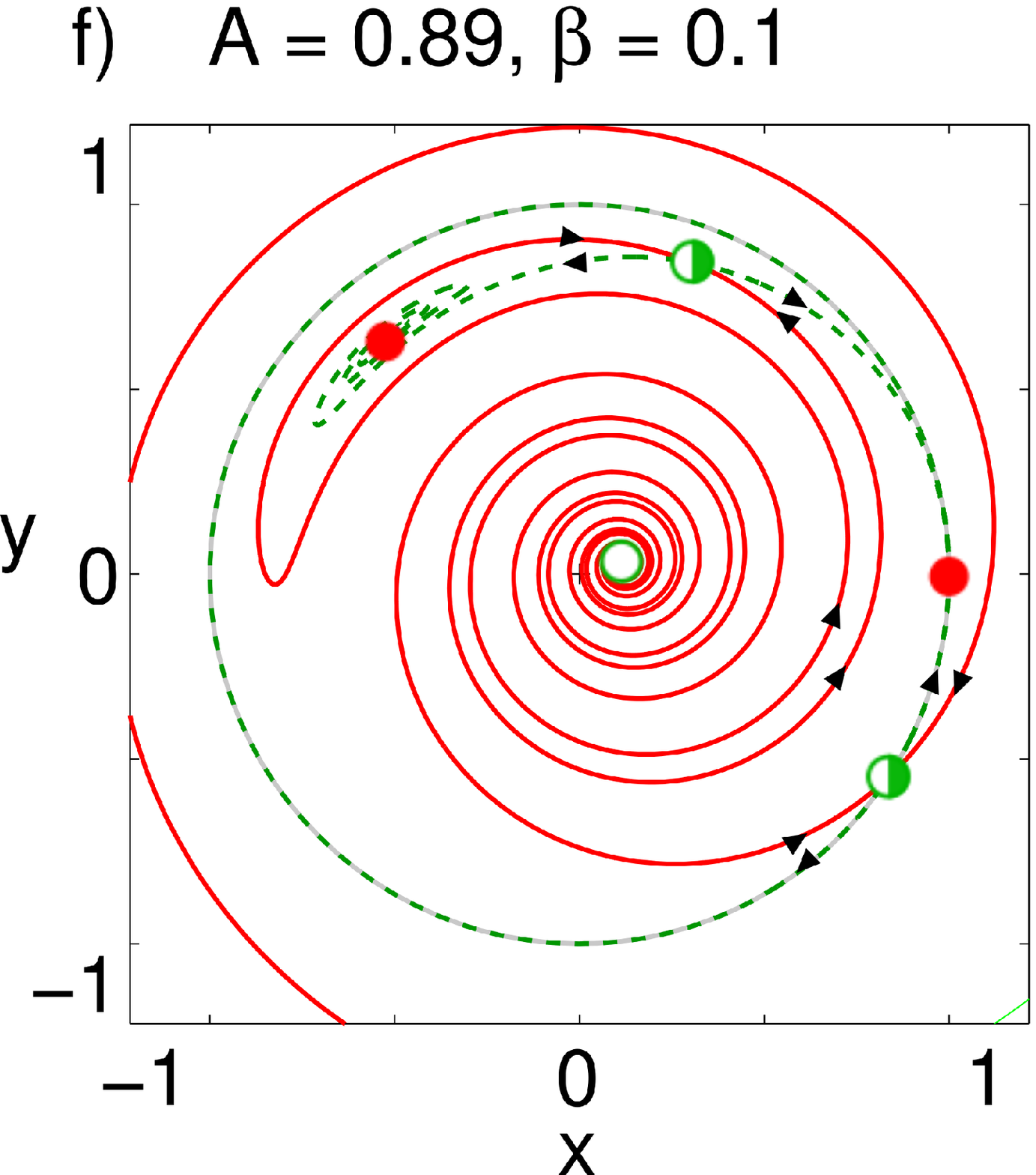} 				
 	\end{center}
\caption{Phase portraits for the $DSD$ chimeras, with increasing values of $A$ at constant $\beta$. Real ($x$) and imaginary ($y$) components of $\overline{a}_{(\sigma)}$ are shown. The unit circle is displayed in gray (the unstable green dashed manifold of the saddle coincides with it). Stable and unstable fixed points are shown as solid (red) and half-filled (green) circles, respectively. Limit cycles are emphasized in blue color. Stable and unstable manifolds are shown as red solid and green dashed trajectories, respectively. The point in $(\rho,\psi)=(1,0)$ is a nodal sink. The position of the saddle depends on $\beta$ and moves in clockwise direction with growing values of $\beta$. 
}
\label{fig:pp_DSD_hybridc1}
\end{figure*}

Unfortunately, the equations for the $SDS$ and $DSD$ states cannot be solved in closed form. Before we get deeper into the matter of analyzing these states, let us get a quick intuition of their behavior by inspecting the phase planes of their corresponding equations, which will guide us in the subsequent analysis. Their phase portraits  shown in Fig. \ref{fig:pp_SDS_hybridc1} and Fig. \ref{fig:pp_DSD_hybridc1} represent a sweep in parameter space with increasing values of $A$ while keeping the value of $\beta=0.1$ constant.

\begin{figure}[ht]
	\begin{center}
		\includegraphics[width=.45\textwidth]{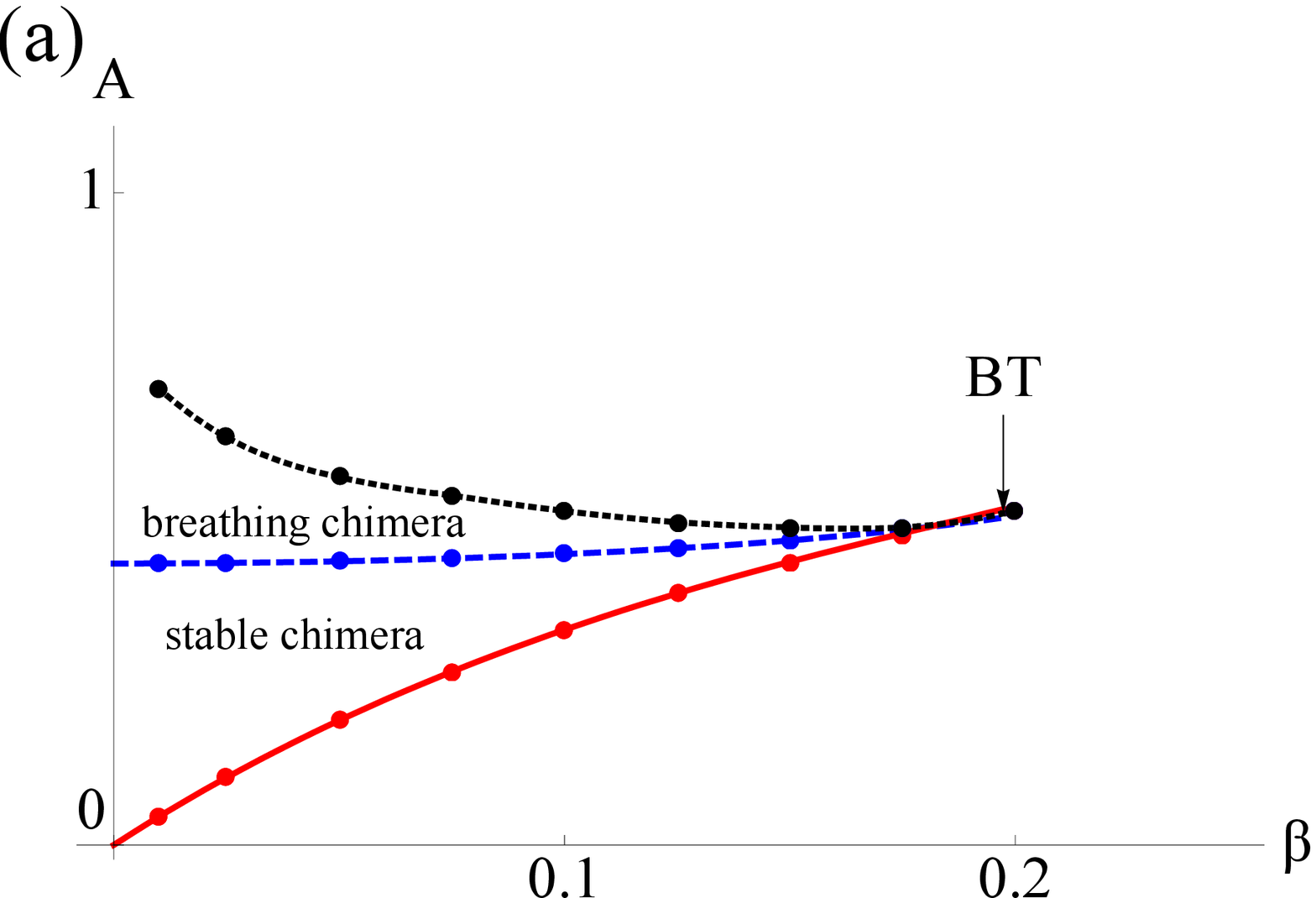} \\
		\includegraphics[width=.45\textwidth]{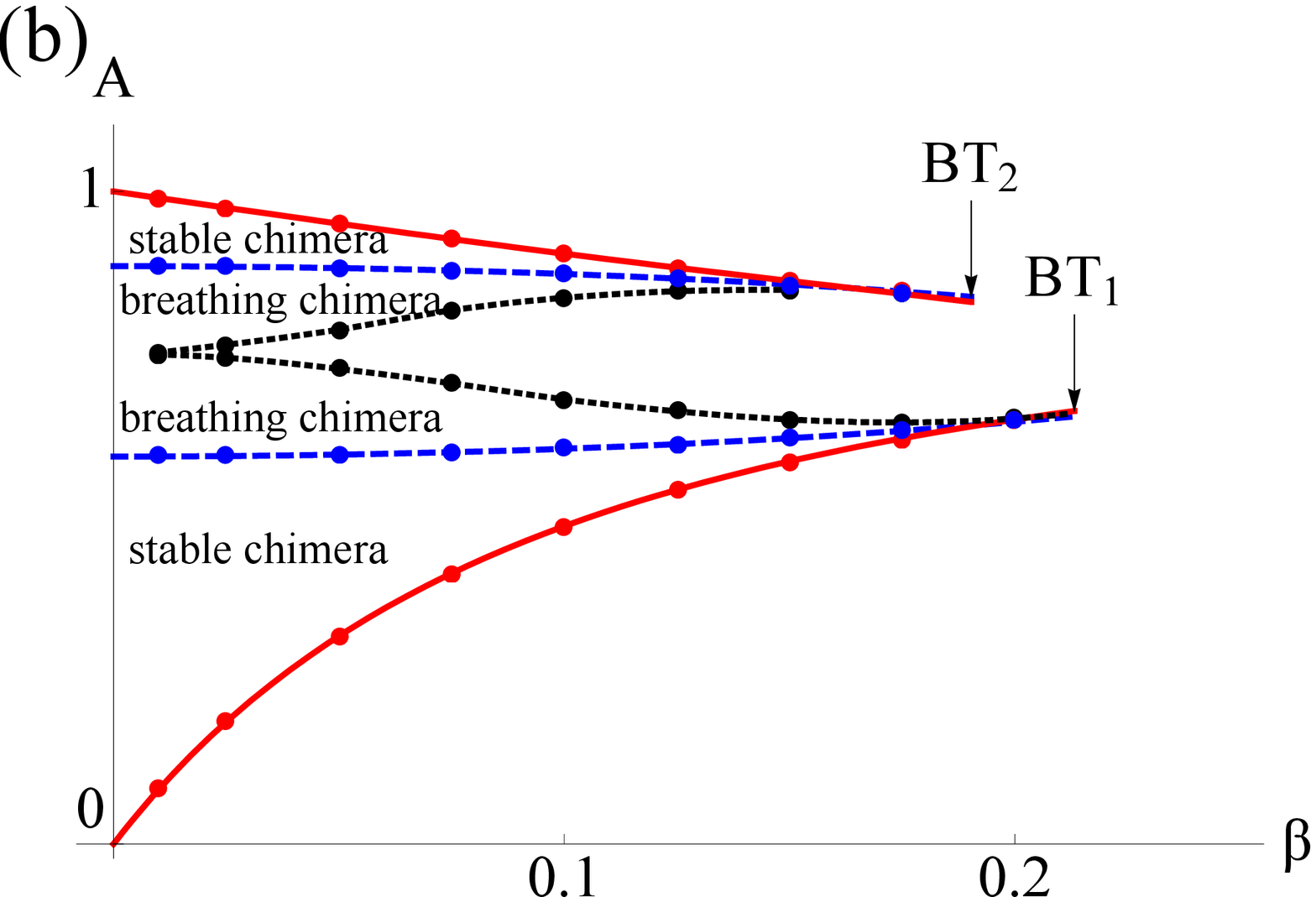} 
	\end{center}
\caption{Bifurcation diagram for the $SDS$ (a) and $DSD$ (b) chimera. The displayed curves are: the saddle-node curve (solid red), the Hopf curve (dashed blue), and the homoclinic curve (dotted black). Dots mark the bifurcation points obtained by inspection of the phase plane. The homoclinc curve (dotted) is an interpolation based on these points, whereas all the solid curves were obtained analytically. }
\label{fig:BD_SDS_hybridc1}
\end{figure}

Consider first the $SDS$ symmetry. For small values of $A$ (close to global coupling), we only observe a nodal sink and source on the unit circle; these points correspond to the in-phase $SSS$ solutions (Fig. \ref{fig:pp_SDS_hybridc1}(a)).  Increasing $A$ further, a saddle-node pair is born very close to the unit circle. For larger $A$, the node moves closer to the origin, implying that the order of the desynchronized population decreases and the chimera state becomes more pronounced (if we instead increase the values of $\beta$, this critical point starts to move clockwise while getting closer to the origin). The node has become a stable spiral (Fig. \ref{fig:pp_SDS_hybridc1}(b)), and at a critical value of $A$, it loses stability through a Hopf bifurcation and a limit cycle is born (Fig. \ref{fig:pp_SDS_hybridc1}(c)). The amplitude of the order parameter $\rho$ of the drifting population starts to oscillate, and is therefore called a \emph{breathing chimera}. As we raise the value of $A$ more, the limit cycle gains in amplitude until it collides with the saddle: the limit cycle is destroyed in a homoclinic bifurcation (Fig. \ref{fig:pp_SDS_hybridc1}(d)). The resulting bifurcation diagram is shown in Fig. \ref{fig:BD_SDS_hybridc1} (a). The saddle-node, Hopf and homoclinic bifurcation curves all intersect in a Takens-Bogdanov point with codimension two.

For the $DSD$ symmetry, we observe a scenario that is qualitatively similar to the previous case. However, surprisingly, the whole scenario appears a second time in the upper part of the parameter plane, but now in reversed order, as shown in Fig. \ref{fig:BD_SDS_hybridc1} (b). We again increase $A$ while keeping $\beta$ constant, as shown in Fig. \ref{fig:pp_DSD_hybridc1}. For all parameter values, we find two synchronized $SSS$ solutions on the unit circle: one is a nodal sink in $(\rho,\psi)=(1,0)$, but what in the $SDS$ case before was a nodal source, is now a saddle. Also notice that a new fixed point has appeared in the left half of the unit circle in the form of a spiral source (Fig. \ref{fig:pp_DSD_hybridc1}(a)). This is the second, currently unstable, $DSD$ chimera seen in the upper half of the bifurcation diagram. As $A$ increases and the coupling becomes more local, a saddle-node pair is born in the right half of the unit circle (Fig. \ref{fig:pp_DSD_hybridc1}(b)); again, its node then becomes a spiral and loses stability as the coupling strength becomes more local, and the resulting limit cycle (Fig. \ref{fig:pp_DSD_hybridc1}(c-d)) gets ultimately destroyed in a homoclinic bifurcation (Fig. \ref{fig:pp_DSD_hybridc1}(e)). 
Whereas one chimera has been rendered unstable, we observe that, above a critical $A$, a stable limit cycle has formed around the spiral source on the left half of the circle:  the twin of the $DSD$ chimera in its breathing mode has emerged (Fig. \ref{fig:pp_DSD_hybridc1}(e)).  From here, the bifurcations happen in reversed order, the source becomes first a spiral node (Fig. \ref{fig:pp_DSD_hybridc1}(f)), i.e. a stable chimera, which is eventually annihilated in a saddle-node bifurcation. The resulting bifurcation diagram is seen in Fig. \ref{fig:BD_SDS_hybridc1} (b).

\subsection{Calculation of bifurcation curves} 
In order to calculate the saddle-node and Hopf curves of the $SDS$ solutions, we must linearize (\ref{eq:SDShybrid}) around the appropriate fixed point. This task amounts to solving the fixed point equations implied by Eqs. (\ref{eq:SDShybrid}) and (\ref{eq:DSDhybrid}) simultaneously with the saddle node condition, 
\begin{eqnarray*}
 \det{(J)}&=&0,\
\end{eqnarray*}
or with the Hopf condition,
\begin{eqnarray*}
 \textrm{tr}{(J)}&=&0 \quad  \textrm{and} \quad \det{(J)}>0,\
\end{eqnarray*}
where $J$ denotes the Jacobian of (\ref{eq:SDShybrid}) or (\ref{eq:DSDhybrid}), respectively. For the $SDS$ symmetry, we have
\begin{eqnarray*}
	J_{11}&=&\frac{1}{2}\left[(1-3\rho^2)\sin{\beta}-4(1-A)\rho\sin{(\beta+\psi)}\right],\\
	J_{12}&=&(1-A)(1-\rho^2)\cos{(\beta+\psi)}, \\\nonumber
	J_{21}&=&\frac{1}{\rho^2}\left[\cos{\beta}(\rho^3-(1-A)\cos{\psi})\right.\\
	      &&-\left.\sin{\beta}\sin{\psi}(2\rho^2-1)(1-A)\right],\\
	 J_{22}&=&\frac{A-1}{\rho}\left[(1+2\rho^2)\cos{\psi}\sin{\beta}+\cos{\beta}\sin{\psi}\right].\
\end{eqnarray*}
The fixed point condition implied by (\ref{eq:SDShybrid}) yields the nontrivial solution
\begin{eqnarray}
 	\rho &=& 2 (A-1) \csc{\beta} \sin{(\beta + \psi)}.\
\end{eqnarray}
Substitution of this expression into the fixed point condition for $\psi$  results in
\begin{eqnarray}\label{eq:SDSpsidoteqn}\nonumber
 0&=& (A-2) \cos{\beta} +  \frac{1}{2} \csc{(\beta + \psi)} \sin{\psi} \\&+& (1-A)^2  
	\csc{\beta}\left[\sin{2 \beta} +  2(\cot{\beta} \sin^2{\psi} + \sin{2 \psi})\right].\
\end{eqnarray}
Unfortunately, this equation cannot be solved in closed form for $\psi$, which in turn would allow us to express $A$ in terms of $\beta$.  We settle therefore for a series approach in $A$ and $\psi$, as follows:
\begin{eqnarray}
	\psi &=& \sum_{k=0}^N\, p_k\,\beta^k + {O}(\beta^{N+1})\\
	A &=& \sum_{k=0}^N \,a_k\,\beta^k+ {O}(\beta^{N+1}).\
\end{eqnarray}
We substitute these two expressions into fixed point equation (\ref{eq:SDSpsidoteqn}) and the saddle node condition, and solve the resulting equations for each power of $\beta$. This leads to the following expression for the saddle node curve:
\begin{eqnarray}\label{eq:SDS_SN_pert_c1}\nonumber
A_{SN}(\beta) &=& \frac{9}{2}\,\beta - \frac{63}{4}\,\beta^2 + \frac{195}{4}\,\beta^3 - \frac{2355}{16}\,\beta^4 \\\nonumber
&&+ \frac{35283}{80}\,\beta^5 - \frac{210247}{160}\,\beta^6 + \frac{872617}{224}\,\beta^7\\\nonumber
&&- \frac{2949379}{256}\,\beta^8 + \frac{2744116261}{80640}\,\beta^9+{O}(\beta^{10}).\\
\end{eqnarray}
Using the  Hopf condition and proceeding in the same way we also find the Hopf curve, approximated by
\begin{eqnarray}\label{eq:SDS_Hopf_pert_c1}\nonumber
 	A_H(\beta) &=& 0.447153+ 1.34639\, \beta^2 + 8.34371\, \beta^4 +{O}(\beta^{6}).\\
\end{eqnarray}
The  Takens-Bogdanov point is determined by numerically solving the saddle-node, Hopf and fixed point conditions simultaneously. It is located at 
\begin{eqnarray}
(\beta,A)_{SDS}=(0.1974, 0.5092).\ 
\end{eqnarray}

The bifurcation curves for the $DSD$ chimeras are obtained in an analogous procedure. Solving the equations expanded in series is now a bit trickier due to the coexistence of two branches. For brevity, we shall only summarize our findings. The two saddle node curves are approximated by
\begin{eqnarray}\label{eq:DSD_SN_pert_c1}\nonumber
A_{SN,1}(\beta) &=&  9 \,\beta - 72 \,\beta^2 + \frac{1059}{2}\, \beta^3 - 3855 \,\beta^4 \\\nonumber
&&+ \frac{1130943}{40}\, \beta^5 - \frac{1039276}{5} \,\beta^6 \\\nonumber
&&+ \frac{854234093}{560} \beta^7- \frac{78311783}{7}\,\beta^8\\
&&+ \frac{3309788681161}{40320} \,\beta^9+{O}(\beta^{10}),\\\nonumber
	A_{SN,2}(\beta) &=& 1 - \beta + \frac{31}{6}\, \beta^3 - \,\beta^4 - \frac{6421}{120} \,\beta^5 +{O}(\beta^{6}).\\
\end{eqnarray}
Unfortunately, the series approach for the lower second branch converges extremely slowly and doesn't match all the way to the Takens-Bogdanov point (even going to this high order doesn't help!). However the effort to find the series coefficients was not all in vain: a Pad{\'{e}} approximant based on the above power series at order three does an excellent job in matching the data points retrieved from the examination of the phase portraits. We have
\begin{eqnarray}\label{eq:DSD_SN2_Pade}\nonumber
A_{SN,2}(\beta) &\approx& \frac{9 \,\beta + \frac{1521018}{301403}\, \beta^2 + \frac{287446827}{6028060}\, \beta^3}
{1 + \frac{2580226}{301403}\, \beta + \frac{90123833}{6028060} \,\beta^2 + \frac{200207828}{4521045}\, \beta^3} .\\
\end{eqnarray}
Finally, the Hopf curves are approximated by
\begin{eqnarray}\label{eq:DSD_Hopf_pert_c1}\nonumber
 	A_{H,1}(\beta) &=& 0.593737+ 1.14491\,\beta^2 \\
			&&+ 4.55308\,\beta^4 +{O}(\beta^{6}),\\\nonumber
	A_{H,2}(\beta) &=& 0.885408- 1.15074\,\beta^2  \\
			&&- 3.96289\,\beta^4 +{O}(\beta^{6}).\
\end{eqnarray}
The Takens-Bogdanov points are located at
\begin{eqnarray}
	(\beta,A)_{DSD,1}&=&(0.2132, 0.6615)\\
	(\beta,A)_{DSD,2}&=&(0.1903, 0.8359).\
\end{eqnarray}

For all symmetries, we found an excellent agreement of our perturbative results with the bifurcation points obtained from inspection of the phase portraits. We did not derive an analytical expression for the homoclinic bifurcation curves; the curves shown in the Figs. \ref{fig:BD_SDS_hybridc1} (a) and \ref{fig:BD_SDS_hybridc1} (b) are based on data points obtained from inspecting the phase portraits while varying the parameters.

\section{Numerical simulations}\label{sec:numericalsimulations}
We have obtained analytical results describing the dynamics of our triangular network of oscillator populations, using two different reductions: firstly, we reduced the governing Eqs. (\ref{eq:goveqn}) using the Ott Antonsen method. 
And secondly, we have assumed that the system attains certain symmetry states that allow for chimera states; these symmetries need not be transversely stable to perturbations off the symmetry manifold. 
Thus, the equations obtained from these reductions may not necessarily account for the complete dynamics of the governing equations, and we checked if our analytical results agree with numerical simulations of the governing equations. 
We did this with a finite, but what may be considered a sufficiently large oscillator population ($N^{(\sigma)=40}$). 

We used two different methods to generate initial conditions that would lead to the appearance of chimera states:
Firstly, for the phases of the desynchronized populations we used an initial condition in the shape of a bump, specifically, a Gaussian distribution in the shape of $\phi_i^{\sigma,d}\sim\exp{(-\gamma(i/N_{\sigma}-1/2)^2)}$ with an appropriately chosen decay rate $\gamma$. 
The second method was chosen with the intention to place the system right on the Ott-Antonsen (OA) manifold. This was accomplished by generating a phase distribution that is consistent with the Poisson kernel (\ref{eq:poissonkernel}). For both methods, the phases for the synchronized populations are given by $\phi_i^{\sigma,s}=0$.
To achieve this, we solved the $SDS$ and $DSD$ Eqs. (\ref{eq:SDShybrid}) and (\ref{eq:DSDhybrid}), respectively, for fixed points $(\rho,\psi)$. 
This in turn, enables us to compute the Poisson distribution $f^{\sigma}(\theta,t)$ defined by (\ref{eq:poissonkernel}), by using the definition of the order parameter, $a_{\sigma}=\rho_{\sigma}e^{-i \phi_{\sigma}}$.
Because this function defines the probability with which oscillators populate a certain phase, we may use its inverse cumulative distribution function to construct from it a set of phases that is consistent with the OA-manifold. The system should remain close to the OA-manifold, because of its invariance. Unless mentioned otherwise, we used this latter method.

We first confirmed that the unreduced system would exhibit all types of chimeras predicted by our analysis, and that they would correspond to stable states, for various points in parameter space. 
These states were observed with both $N_{\sigma}=20$ and $N_{\sigma}=200$ oscillators per population. 
The observed behavior is that the system first goes through a tiny transient and reach an attracting state which was confirmed to be stable even for long computation times. 
The transient may be explained by the fact that the system, due to its finite size, can only be approximately on to the OA-manifold (or a member of the OA-family, as explained in the Discussion).

\begin{figure}[ht]
	\begin{center}
		\includegraphics[width=0.48\textwidth]{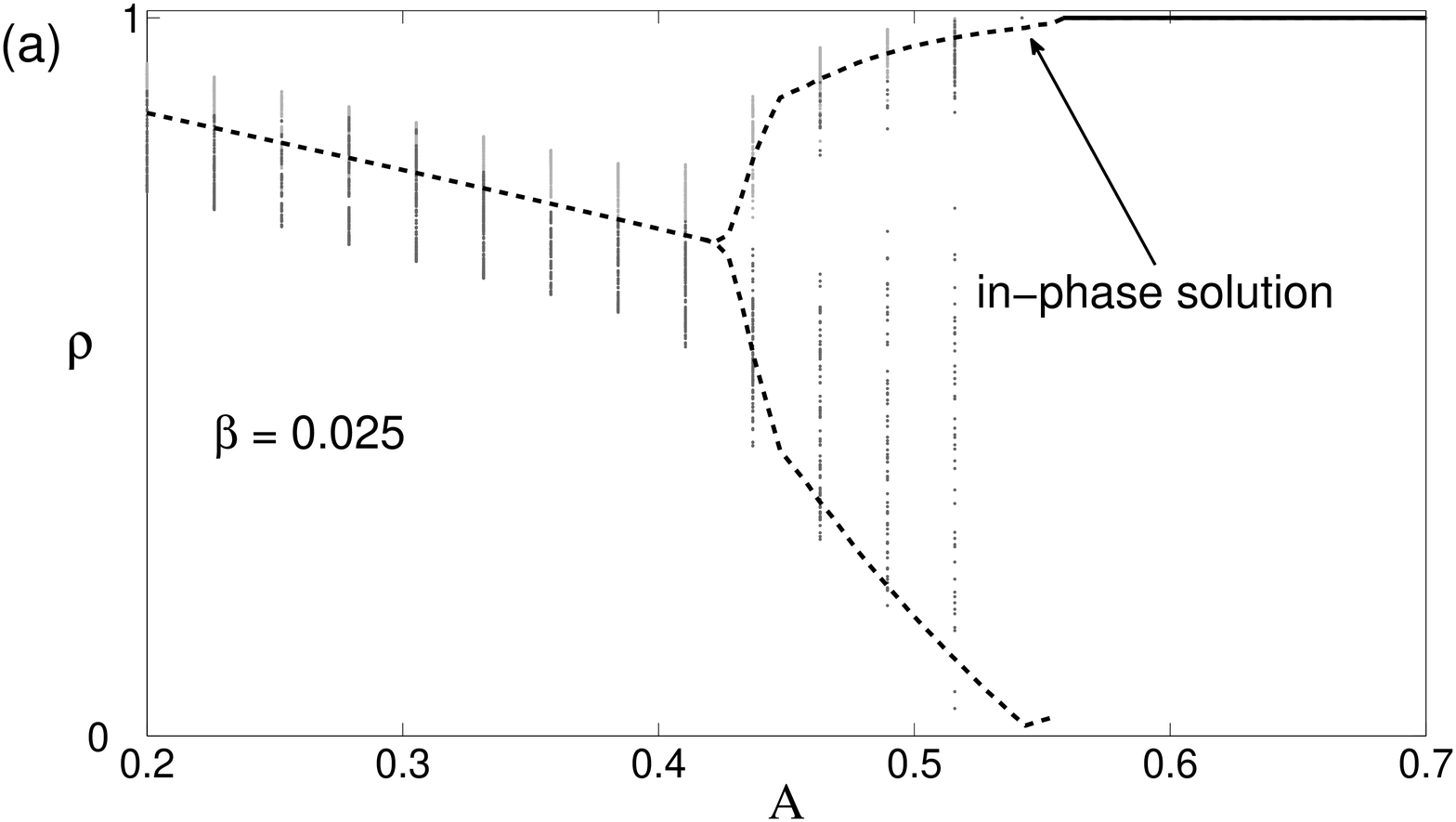}\\
		\includegraphics[width=0.48\textwidth]{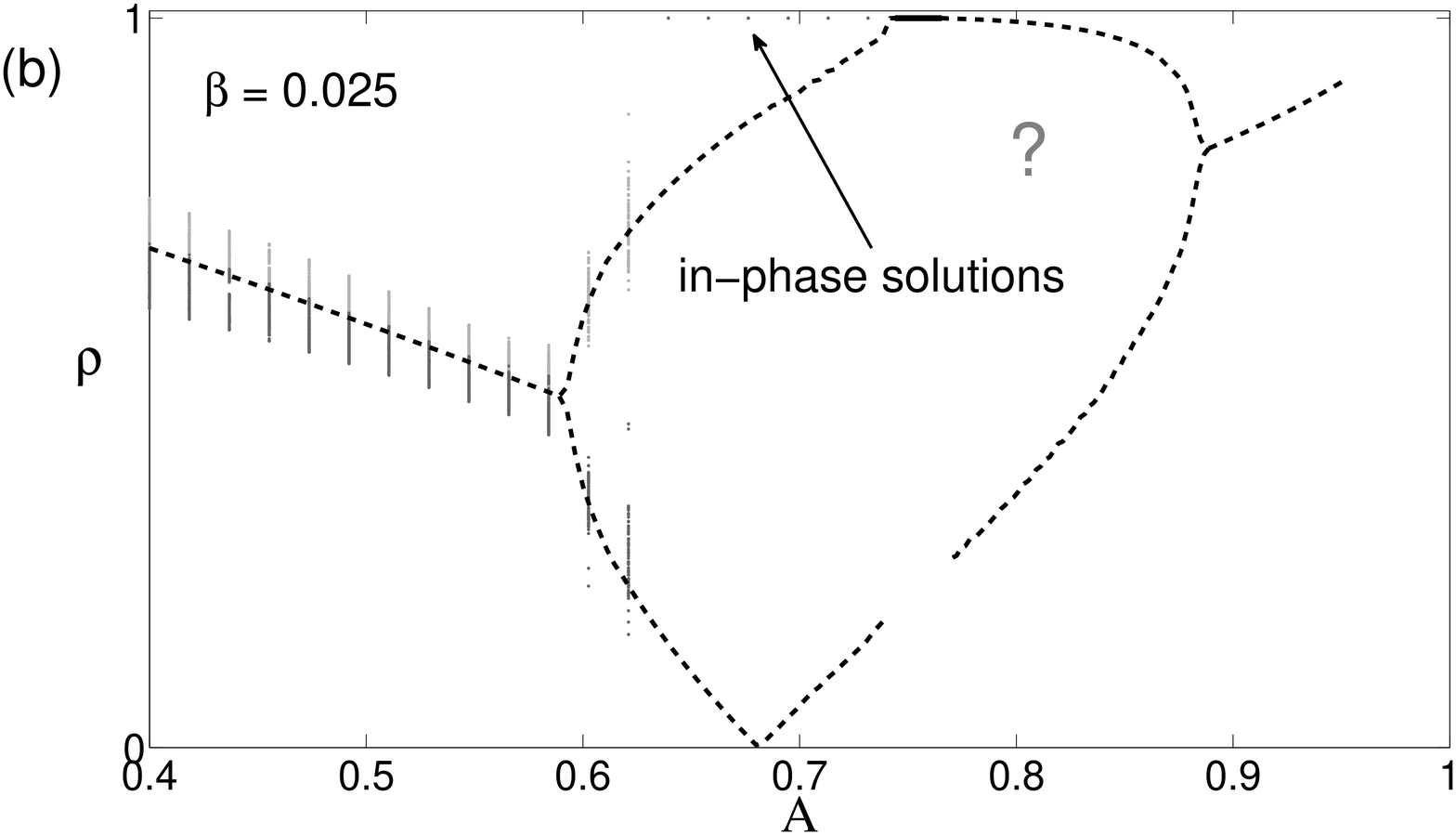}\
	\end{center}
	\caption{Bifurcation diagram obtained from numerical simulation, for states $SDS$, shown in (a), and $DSD$ in (b). Light dots and dark circles correspond to local maxima and minima, respectively, that are detected by the algorithm. The dashed curve represents the analytical result for the continuum case ($N\rightarrow \infty$). The computations were performed with $N_{\sigma}=40$ oscillators per population for a simulation time of $T=100$. The question mark indicates that we could not conclusively confirm the existence of the second DSD state. (The kink in the lower left dashed branch is an artefact from the limit cycle reaching into the left hand side quadrants, as $\rho$ is not measured relative to the limit cycle center but to the origin.)}
	\label{fig:SimuBD}
\end{figure}

Next we checked if the critical parameter values of saddle node, Hopf and homoclinic bifurcation in the full system would be in accord with the critical values obtained from our theory, see Figs. \ref{fig:SimuBD} (a) and (b). In order to do so, we held the value of $\beta$ fixed and continued a solution through $20$ increasing values of $A$. To initiate the continuation we used an initial condition consistent with the OA-manifold.

In Fig. \ref{fig:SimuBD}, we show fixed points and oscillation amplitudes of the order parameter, obtained analytically from (\ref{eq:SDShybrid}) and (\ref{eq:DSDhybrid}), as  dashed curves. 
Our simulation results are superposed, and were obtained as follows:
To remove transient effects from our analysis, we only considered the last $2/5$ of the computed time series. 
Instead of only detecting the global maximum and minimum of the series, we detected local maxima, shown as light gray dots, and minima, shown as dark gray dots. We chose to do this, because it would allows us to see the traces of new appearing periods, that potentially may occur in the unreduced, highly dimensional system.
However, finite size fluctuations also cause small oscillations; this is the reason we see some small amount of blurring in the data (For similar reasons, maxima and minima on the stable branch do not coincide.) 

Despite these undesired effects, we can clearly demonstrate the onset of the Hopf and the homoclinic bifurcations in the simulation. Consider first the $SDS$ chimera, shown in Fig. \ref{fig:SimuBD} (a). We expect that finite size effects affect the locations of all the bifurcation points. While we would have to compute at a higher resolution for the continuation to actually see this happen for the Hopf bifurcation, it is more apparent for the homoclinic bifurcation: the limit cycle oscillation brings the system periodically close to the saddle point (as seen in the phase plane); therefore, the larger finite size fluctuations are, the more likely the system is to be kicked off the limit cycle and is instead attracted to the nearby all-in-phase $SSS$ state on the unit circle. This is seen in Fig. \ref{fig:SimuBD} (a), where the limit cycle oscillation disappears at a smaller value of $A$ than it does for the analytic result (sold curve indicates $SSS$ state). 
The same behavior is observed for the $DSD$ chimera in Fig. \ref{fig:SimuBD} (b), but much more pronounced: the system snaps to the in-phase state (solid curve) much earlier than expected for a continuous system of oscillators. 
Similarly, we can continue the solutions in reverse direction, and check that the chimera states are annihilated via a saddle node bifurcation.

Whereas we easily managed to show $DSD$ states in the lower half of the parameter plane, our attempts to place the system on the second $DSD$ attractor had little success: for all trials the system would eventually reach the in-phase state. Increasing the number of populations didn't seem to remedy the matter, as one might be led to think from our previous experience; rather, the system would stay obediently in the $DSD$ state for a little while and then suddenly jump into the in-phase state. However, it doesn't seem to be entirely clear whether the second attractor of the $DSD$ state is inherently unstable, or if it is just very hard to stay on that manifold due to a combination of factors, given by finite size effects and the location and shape of the attractor (see phase plane in Fig. \ref{fig:pp_DSD_hybridc1}). Specifically, in the case of a breathing state, the limit cycle is always very close to the invariant field defined by the in-phase state on the unit circle. In conclusion, it is likely that this second $DSD$ attractor is unstable to symmetry breaking perturbations.

\section{Discussion}\label{sec:discussion}
We have investigated a triangular network of populations of sinusoidally, nonlocally coupled oscillators with identical frequencies. Our analysis approximates the practical case of large populations by considering the limit of infinitely many oscillators per population, for which we may use the Ott-Antonsen method to reduce the governing to a finite set of equations \cite{ott2008ldb}. By assuming symmetries compliant with chimera states, we further reduced these equations and studied their emergence in a phase plane analysis. Saddle node and Hopf bifurcation curves are determined using perturbation techniques, and homoclinic bifurcation curves by observation of phase portraits. 
The resulting stability diagram is a variation of the one found by Abrams \emph{et al.} \cite{abrams2008smc}, but is governed by the emergence of two different types of chimeras: one with a desynchronized 'zone' of narrow width ($SDS$), and the other with a desynchronized zone of twice the width ($DSD$). 
$DSD$ chimeras exist in two regions of parameter space, and differ in their mean (difference) phase angle $\psi$; for the $DSD$ state near global coupling (small $A$), $\psi$ remains close to zero, and for the other near local coupling (large $A$) closer to $\pi$. Numerical experiments demonstrate that chimeras persist for small oscillator populations, and that the $SDS$ and the $DSD$ chimera near global coupling are truly stable attractors in the unreduced system described by Eqs.	 (\ref{eq:goveqn}). Finally, we observe for the first time the feature of bistable chimeras.

\subsection{Reduced system and numerical experiments}
Ott and Antonsen \cite{ott2009lte} showed  that the long time dynamics of a large class of phase oscillator systems (including this one) is attracted to the manifold defined by (\ref{eq:poissonkernel}) if the natural frequencies are drawn from peaked distributions. This result has been confirmed in several studies \cite{barreto2008snn,childs2008,laing2009csh,martens2008erk}. 
The situation is somewhat different if we consider the case of identical frequencies. Pikovsky and Rosenblum \cite{pikovsky2008pid}, and Marvel \emph{et al.} \cite{marvel2008iss} have shown independently that the dynamics of each population can be reduced to a flow described by three variables plus constants of motion. 
The unifying picture is that for identical frequencies one has a whole \emph{one-parameter family} of invariant manifolds, including the OA manifold. These manifolds are neutrally stable with regards to perturbations in transverse direction to themselves (such perturbation would result in placing the dynamics into a neighboring manifold). Once the frequencies are spread this family collapses and the OA manifold is left.

Because this degeneracy in the case of identical frequencies might lead to unanticipated results, and because the symmetry manifolds defined by $SDS$ and $DSD$ don't need to be stable to perturbations in transverse directions, we checked that the chimera states are true attractors in the unreduced system, Eqs. (\ref{eq:goveqn}), by numerical simulation. 
The sequence of bifurcations with saddle node, Hopf and homoclinic bifurcations is indeed reproduced in the unreduced system and the bifurcations appear close to the predicted critical values, even though the number of oscillators per population is small. The homoclinic bifurcation makes an exception in the sense that it seems to happen already for small $A$; we have argued that this likely is an artefact for large limit cycles, where the trajectory gets kicked off the orbit because of finite size fluctuations in combination with the appearance of a nearby saddle point. All chimera states, with exception of the $DSD$ state near local coupling, have proved to be true attractors in the unreduced system.

\subsection{Relation to heterogeneous frequency distributions}
A recent study by Laing \cite{laing2009csh} generalizes the problem of 
a network with two oscillator populations investigated by \cite{abrams2008smc} to the case of heterogeneous frequencies.
Laing showed that for this and various other network topologies, the chimera state is robust -- within limits -- to heterogeneity in the intrinsic frequencies of the oscillators.
In particular, he finds that the chimera state remains stable for populations with nearly identical oscillators, that is, with a narrow width of the distribution. The stability results obtained from our analysis should therefore	 be the same as the one obtained for the dynamics of oscillators with almost identical frequencies.

\subsection{Aperiodicity and chaos}
In our setting of the problem with identical oscillators, we were able to find solutions that do not lie on the OA-manifold (to find them, use initial conditions lying off the OA-manifold, e.g. Gaussian 'bumps' or add noise to the OA-compliant initial conditions). Such states may be related to the quasi-periodic states shown to exist by Pikovsky \emph{et al.} \cite{pikovsky2008pid}, but perhaps also to chaotic states. 
We haven't gone further into this question, but have noticed irregular behavior in some of our simulations that look aperiodic, and we think it might be possible to find something like a \emph{chaotic} chimera state.

\subsection{All-in-phase states} 
In the case where all populations are fully synchronized, i.e. $\rho_\sigma=1$, the dynamics of the system effectively becomes the one governed by three coupled oscillators. The stability of such states is analyzed in the Appendix. This case has already been studied in the context of three coupled sinusoidal limit cycle oscillators by Mendelowitz \emph{et al.} \cite{mendelowitz2009dtc} and for relaxation limit cycle oscillators by Bridge \emph{et al.} \cite{bridge2009drt}, who found that two rotating waves, clockwise and counter-clockwise rotation, are possible.

\subsection{Relation to ring of oscillators and network with two oscillator 
populations}
We mention the connection of our study to the work done by Abrams et al. \cite{abrams2008smc}, who studied a similar system, but with only two oscillator populations.  With its two in-phase-locked populations, it is unclear whether the $SDS$ or the $DSD$ state compares best to their -- using our terminology -- $SD$ chimera. Certainly, it can be said they both act like a two population system in disguise. In this sense, the two population system is a degenerate case of our triangular network.

The same authors also studied a continuum of oscillators on a ring \cite{abrams2004csc,abrams2004cro}. In our picture with oscillator populations, this system is approximated by an infinite set of populations arranged on a ring.  
For the continuous ring, only a single chimera is known. Both $SDS$ and $DSD$ chimeras effectively act like a system made of two populations, but differ in their 'width' of drifting populations. In the case of the ring, this width is slightly larger than the synchronized region; from this point of view, the $DSD$ chimera seems to match their state best.

One may ask how the behavior of discrete ring-like systems is affected as we increase the number of populations.
One could  imagine that more and more chimera states get added as we increase the number of populations, constituting competing multistable attractors. But then, what happens as we take this continuum limit? Do they disappear, collapse such that only one of them, specifically, the one discussed in \cite{abrams2004cro}, wins the competition, and dominates over all others and remains stable?  In the case of the triangular network, the $SDS$ and the first $DSD$ chimera are truly stable attractors in the unreduced system, and so the situation stays inconclusive. The creation and annihilation of chimera states with varying numbers of oscillator populations is also of importance if we want to characterize the basins of attractions in complex oscillator networks, an issue pointed out recently by Motter \cite{motter2010}.

\subsection{Multistability and neural networks}
Our numerical experiments show that the existence of chimeras not only exist in the thermodynamic limit of $N\rightarrow \infty$, but also for relatively low numbers of oscillators per population of $N \sim 20$. 
In view of the robustness of chimera states towards heterogeneity \cite{laing2009csh}, one may therefore expect that chimeras also may emerge in biological settings such as neuronal networks. In this case, the  coexistence of chimera attractors would have implications for the study of neuronal networks beyond their similarity to spatially localized "bumps" of neuronal activity mentioned earlier. 
For instance, one may envision that multiple coexisting chimera attractors encode states of associative memory \cite{mendelowitz2009dtc,hoppensteadt1999} or play a role in decision making, where different initial conditions lead to different states. The characterization of basins of attractions for chimeras therefore represents an interesting issue, and represents a topic in the field that is not well explored \cite{motter2010}. 
Alternatively to stable-state dynamics, some neuronal models, based on the mechanism of switching among unstable saddles, have drawn attention lately \cite{aswhinnature2005}. In these models, heteroclinic connections generate characteristic patterns of neural activity and thus represent information. One may conceive oscillator networks with the possibility of switching between chimera states while varying system parameters. Provided that such chimera sequences, encoding different spiking activity patterns, are realizable, they may constitute an interesting field of research.
We note that in this context oscillator populations, or rather neurons, represent redundancy in computing networks, which significantly enhances their reliability, as shown already by von Neumann \cite{neumann1954}. In a recent experimental study Feinerman \emph{et al.} \cite{feinerman2008} build neuronal computational devices and demonstrate how reliability is significantly improved by employing neuronal ensembles for each logic unit, similar to the oscillator populations.

\section{Acknowledgments}
This research was supported in part by NSF Grant No.
DMS-0412757 and CCF 0835706. I would like to thank Steve Strogatz for advice, mentoring and helpful guidance throughout this project, and Fabio Schittler-Neves for useful conversations.

\appendix

\section{Stability of Symmetric States}
\subsection{Existence of Symmetric States $SSS$}
The states we consider here are fully synchronized, i.e. $\rho_{\sigma}=1$ for all $\sigma = 1,2,3$. 
We consider the symmetries where $\phi_1=\phi_3$.
In order to derive a fixed point condition for these states, it is therefore sufficient to consider the equations that are specialized to the symmetries $SDS$ and $DSD$ given by Equations (\ref{eq:SDShybrid}) and (\ref{eq:DSDhybrid}). Applying the fixed point equation to either of them yields the condition
\begin{eqnarray}\label{eq:SSSFP}
 	0&=& (1-A)(\cos{\beta}-\cos{\beta}\cos{\psi}+3\sin{\psi}\sin{\beta}),\
\end{eqnarray}
which is reduced to these cases:
\begin{eqnarray}\label{eq:SSSFP1}
 	A&=&1,\\\label{eq:SSSFP2}
	\beta&=&0 \quad\textrm{ with }\cos{\psi}=1,\;\sin{\psi}\neq 0,\\\label{eq:SSSFP3}
 	\psi &=&0,\\\label{eq:SSSFP4}
 	\cos{\beta}&=&\frac{3\sin{\beta}\sin{\psi}}{\cos{\psi}-1} \textrm{ with } \cos{\psi}\neq 1.\
\end{eqnarray}
The first two conditions are the degenerate cases representing the $A$ and $\beta$ axis. The third condition corresponds to the fixed point $(\rho,\psi)=(1,0)$, and the last to the position of the node that is constrained to move on the unit circle in the phase portraits, see Figs. \ref{fig:pp_SDS_hybridc1} and \ref{fig:pp_DSD_hybridc1}.

\subsection{Stability of $SSS$ States}
The computation of stability of these points is accomplished by computation of the Jacobian of the six dimensional system (\ref{eq:lodimbeta}), with the coordinate system $(\rho_1,\rho_2,\rho_3,\phi_1,\phi_2,\phi_3)$, using a computer algebra system. Applying our symmetry assumptions, we find the eigenvalues
\begin{eqnarray}
 	\lambda_k = \left(\begin{array}{c}
			0\\
 	             	-\sin{\beta}+2(A-1)\sin{(\beta+\psi)}\\
			(A-2)\sin{\beta}+(A-1)\sin{(\beta-\psi)}\\
			(A-2)\sin{\beta}+(A-1)\sin{(\beta-\psi)}\\
			(A-1)(2\sin{\beta}+\sin{(\beta-\psi)})\\
			(A-1)(3\cos{\psi}\sin{\beta}+\cos{\beta}\sin{\psi})\
 	            \end{array}
		\right).
\end{eqnarray}
The first eigenvalue is a manifestation of the rotational invariance of the system (the system only depends on phase differences and is effectively five dimensional). 

We first compute the stability of the trivially symmetric state defined by $\psi=0$. This degenerate state has the eigenvalues 
\begin{eqnarray}
  0 &=& \lambda (\lambda - (2A-3)\sin{\beta})^3(\lambda-(A-1)\sin{\beta})^2\
\end{eqnarray}
If we only consider parameter values $A\in (0,1)$, a sufficient condition for linear stability is given by $\beta\in(0,\pi)$.

The less trivial state with $\psi\neq 0$ must be considered in combination with the fixed point condition (\ref{eq:SSSFP4}). The signs of these eigenvalues were not obtained analytically but rather by graphing the maximal eigenvalue in the $(\beta,A)$-plane.  It turns out that this state  has at least one positive eigenvalue except for the degenerate case where $\beta=0,\pi$, and is thus always a saddle. This result is consistent with the behavior observed by inspection of the phase portraits of Eqs. (\ref{eq:SDShybrid}) and (\ref{eq:DSDhybrid}) (in the case of $SDS$ symmetry, the nontrivially symmetric state changes stability at $\beta=\pi$, but this holds only within the $SDS$-symmetry manifold, and has nothing to do with stability in the six (or five, for that matter) dimensional space.).

It is possible to obtain a similar stability result for the general case of a network with arbitrarily many populations $N$, for the trivially symmetric point satisfying $\rho_{\sigma}=1$ and $\phi_1=\phi_2=...=\phi_N$, using Gershgorin's circle theorem. In this general setting, the point also becomes linearly unstable as $\beta>\pi$ (provided that all row sums of the coupling matrix are strictly positive). Unfortunately, no similar result was obtained for the remaining  $N-2$ fixed points that may occur on the unit-sphere related to the general problem. The calculation is tedious and will not be represented here.

\bibliography{PhDThesis}

\end{document}